\documentclass[aps,12pt,preprintnumbers,nofootinbib,superscriptaddress]{revtex4}
\usepackage{graphicx}
\usepackage{amssymb,amsmath}
\begin{document}

\title{\Large{\bf  Annihilation Process of Quantum Vortices in Dissipative Gross-Pitaevskii Equation Model } }

\author{Shanquan Lan}
\email[ ]{shanquanlan@126.com}
\affiliation{%
Department of Physics, Lingnan Normal University, Zhanjiang, 524048, Guangdong, China}
\author{Weiru Chen}
\affiliation{%
Department of Physics, Lingnan Normal University, Zhanjiang, 524048, Guangdong, China}
\author{Guoliang Zhang}
\affiliation{%
Department of Physics, Lingnan Normal University, Zhanjiang, 524048, Guangdong, China}
\author{Xiaoying Liang}
\affiliation{%
Department of Physics, Lingnan Normal University, Zhanjiang, 524048, Guangdong, China}
\author{Jiexiang Chen}
\affiliation{%
Department of Physics, Lingnan Normal University, Zhanjiang, 524048, Guangdong, China}
\author{Xiyi Liu}
\affiliation{%
Department of Physics, Lingnan Normal University, Zhanjiang, 524048, Guangdong, China}

\date{\today}

\begin{abstract}
In two dimensional superfluid, annihilation processes of vortices are investigated by numerical simulation within the dissipative Gross-Pitaevskii equation (GPE) model. First, quantum vortex solution is obtained and its fitting function is found. Second, the simulation show that positive and negative vortices accelerate in both x,y directions, until they annihilate into a soliton and then a crescent-shaped shock wave. The processes are found to be controlled by the dissipative parameter and the general Magnus force.  For the behavior of separation distance between vortices $d(t)$, an universal scaling exponent 1/2 is found which is same with the three dimensional cases. Third, system's energy is surprisingly  found to be determined by system's configuration and their relation are obtained. Then we derive the general Magnus force which decreases with the increases of $d(t)$ for large  $d(t)$ and increases with the increases of $d(t)$ for small $d(t)$.

\end{abstract}
\pacs{}
\maketitle
\newpage

\date{\today}

\section{Introduction}

Quantum vortices are topological defects in the order parameter which describes superfluid. In turbulent superfluid, the reconnection of vortex lines for three dimensional cases, or the annihilation of clockwise rotating vortex and counterclockwise rotating vortex for two dimensional cases are important phenomena. These processes reduce the topological defects and randomise the velocity field. What's more, they are essential way of energy dissipation and redistribution. Therefore, this topic attracts a lot attention and is intensively studied. In recent years, H$\ddot{a}$nninen studies the spectrum of Kelvin waves from vortex reconnection in superfluid helium\cite{hanninen2015}. Serafini et al. report experimental and numerical observations of double reconnections, rebounds and ejections in trapped atomic Bose-Einstein condensates\cite{serafini2017}. Hannay calculates the vortex reconnection rate, and loop birth rate for a random wavefield\cite{hannay2017}. While we focus on the behavior of the minimum distance between vortex lines $d(t)$ during the reconnection process. Based on Schwarz's vortex filament model\cite{1988prbschwarz}, de Waele and Aarts firstly report that
\begin{equation}
  d(t)=(\kappa/2\pi)^{1/2}\sqrt{t_{0}-t},
\end{equation}
where $\kappa$ is the circulation quantum and $t_{0}$ is the reconnection time\cite{1994prlwaele}. This scaling exponent 1/2 is later confirmed in Helium experiment\cite{2008PNASbewley,2008pdnppaoletti}, and by an approximate analytic solution in the GPE model\cite{2003jltsergey}. However, many numerical simulations show that the formula of $d(t)$ should be modified as
\begin{equation}
   \delta(t)=A_{1}(t_{0}-t)^{A_{2}},
\end{equation}
or
\begin{equation}
   \delta(t)=B_{1}(t_{0}-t)^{1/2}[1+B_{2}(t_{0}-t)]
\end{equation}
with $A_{2}$ varying around 1/2 and small $B_{2}$ \cite{2012prbbaggaley,2011jbaggaley,2012pfzuccher,2014praallen,2017prfalberto,2019lan}. Galantucci et al. report two universal scaling laws 1/2 and 1, the former arising from the mutual interaction of the reconnecting strands and the latter arising when extrinsic factors drive the individual vortices\cite{2019galantucci}.

The above researches are about three dimensional cases. The two dimensional cases should also be important and interesting. In experiment, there are oblate Bose-Einstein condensate where the vortices are clockwise rotating or counterclockwise rotating. They are simple than vortex lines in three dimension. During their annihilation process, does the distance $d(t)$ has similar behavior and scaling exponent? In this paper, we try to answer this question by numerical simulation within the dissipative GPE model. Additionally, we are going to reveal the details and the energy behavior of the process.

The outline of the paper goes as follows. In section \ref{sec2} we introduce the superfluid model and relevant numerics. In section \ref{sec3} we report numerical results and new findings, e.g. vortex solution in cylindrical coordinate, the detail of vortices' annihilation process, energy behavior. In the last section, we end with a conclusion and discussion.

\section{Two Dimensional Superfluid Model and Relevant Numerics}
\label{sec2}

The two dimensional superfluid model in this paper is the dissipative Gross-Pitaevskii equation (GPE)\cite{epgross1963,pitaevskii1959,lppitaevskii1961,choi1998,tsubota2002}, which is also known as the nonlinear Schr\"{o}dinger equation
\begin{equation}
  (i-\eta)\hbar\frac{\partial\psi}{\partial t}=-\frac{\hbar^{2}}{2m}\nabla^{2}\psi+V_{0}|\psi|^{2}\psi-\mu\psi,
\end{equation}
where $\eta$ is the dissipative parameter which is added by hand, $\psi(\vec{x},t)$ is the wave function for N bosons of mass m, $V_{0}$ is a coupling constant denotes the interaction between the bosons, $\mu$ is the chemical potential.

For the numerical simulation of system's dynamics, dimensionless GPE is practical
\begin{equation}
  (i-\eta)\frac{\partial\psi}{\partial t}=-\nabla^{2}\psi+|\psi|^{2}\psi-\psi.
\end{equation}
This is achieved by applying the rescaling transformation
\begin{eqnarray}
  t\rightarrow \frac{\hbar}{\mu}t,\boldsymbol{x}\rightarrow \frac{\hbar}{\sqrt{2m\mu}}\boldsymbol{x},\psi\rightarrow\sqrt{\frac{\mu}{V_{0}}}\psi.
\end{eqnarray}

The superfluid velocity is
\begin{eqnarray}
  \boldsymbol{u}=\frac{\boldsymbol{J}}{|\psi|^{2}},\,\,\,\boldsymbol{J}=i(\psi\boldsymbol{\partial}\psi^{*}-\psi^{*}\boldsymbol{\partial}\psi)
\end{eqnarray}
and the vortex winding number is
\begin{eqnarray}
  \omega=\frac{1}{2\pi}\oint_{c}d\boldsymbol{x}\cdot\boldsymbol{u},
\end{eqnarray}
here c is a counterclockwise oriented path surrounding a single vortex. From now on, a vortex with $\omega>0$ will be called positive vortex and a vortex with $\omega<0$ will be called negative vortex.

To investigate the dynamical evolution process of a vortex pair, we consider an uniform superfluid $\psi 1(x,y)=1$ in a $60\times60$ square box with coordinate system,$x\in(-30,30)$ and $y\in (-30,30)$. After a positive vortex with $\omega=1$ is located at point(0,10), and a negative vortex with $\omega=-1$ is located at point(0,-10), the uniform superfluid become $\psi 2(x,y)$. Next, $\psi 2(x,y)$ is multiplied with a function $e^{i\varphi(x,y)}$ which is carefully constructed to make the system periodic. Thus the initial superfluid $\psi (t=0,x,y)$ is constructed. Then the system evolves governed by the dissipative GPE. In $x,y$ direction, the pseudo-spectral method is used to represent $\psi$ with 361 Fourier modes. In $t$ direction, the fourth-order Runge-Kutta method is used to evolve GPE  with time step $\delta t=0.002$. For a detail review of these numerical methods, one can refer to Ref.\cite{chesler2013holographic,du2014holographic,2016jheplan,2016cqgguo}.

\section{Results}
\label{sec3}

\subsection{vortex solution}

As quantum vortex has cylindrical symmetry, we will work in the cylindrical coordinate $(\rho,\theta)$. The complex function $\psi(x,y)$ can be rewritten as $\phi(\rho)e^{i\omega\theta}$ where $\omega$ is the vortex winding number. Here $\omega$ have to be integers to guarantee $\psi(x,y)$ is a single-valued function. The GPE is rewritten as
\begin{equation}
  \frac{\partial^{2}\phi(\rho)}{\partial \rho^{2}}+\frac{\partial \phi(\rho)}{\rho\partial\rho}-\frac{\omega^{2}}{\rho^{2}}\phi(\rho)-\phi^{3}(\rho)+\phi(\rho)=0.
\end{equation}
\begin{figure}
		\centering
		\includegraphics[scale=0.6]{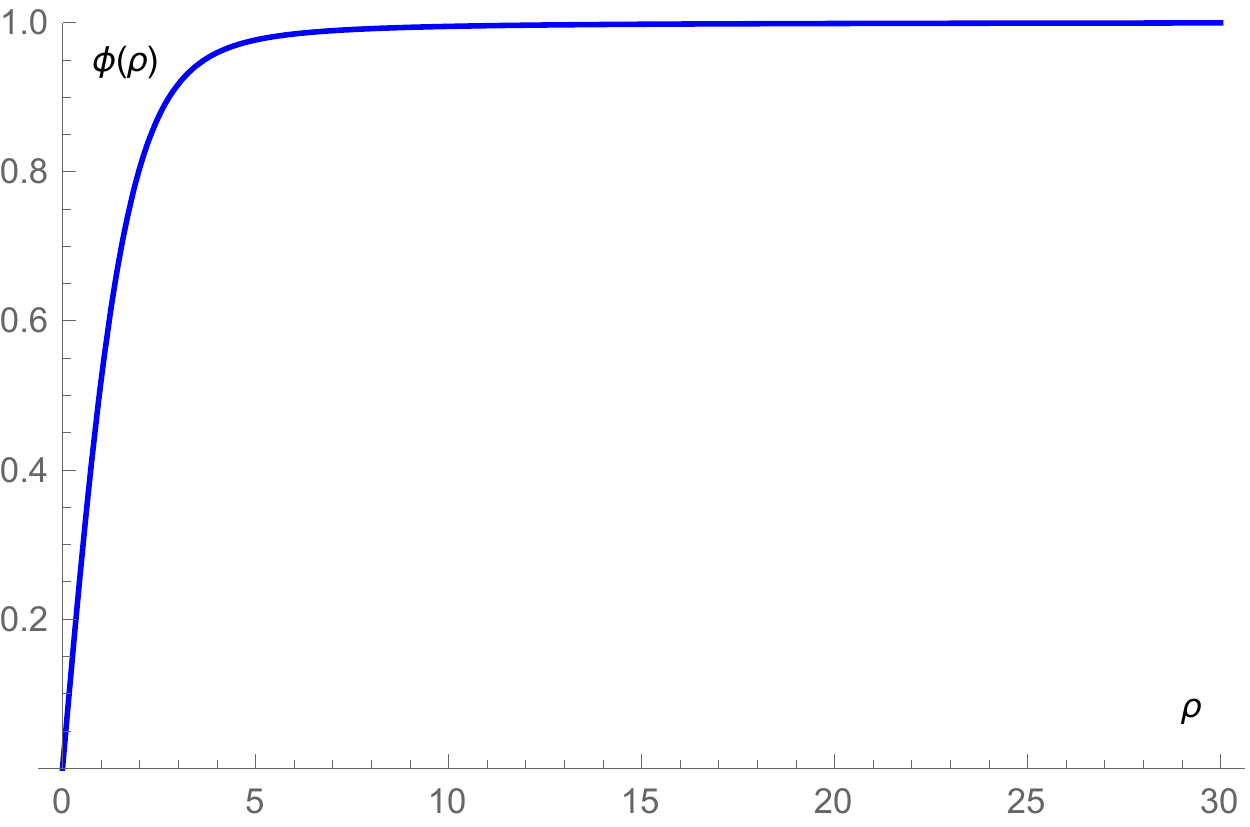}
	\caption{Vortex configuration for $\omega=\pm 1$.}
	\label{svortex}
\end{figure}

From the third term of the above equation, it can be clearly seen that the positive vortex and negative vortices have the same configuration. The only difference is that one has counterclockwise rotation while the other has clockwise rotation. The above equation can be solved numerically for $\rho \in (0,L)$. At $\rho=0$, we choose Dirichlet boundary condition $\phi(0)=0$. At $\rho=L$, we choose Dirichlet boundary condition $\phi(L)=1$ or Neumann boundary condition $\phi'(L)=0$ which give same solution. Set $L=30$, the vortex configuration for $\omega=\pm 1$ is shown in Fig.\ref{svortex}. The vortex solution is approximately fitted by
\begin{equation}
  \phi(\rho)=tanh(0.557\rho),
\end{equation}
and well fitted by
\begin{equation}
  \phi(\rho)=0.971tanh(0.587\rho)+0.00270\rho-0.0000621\rho^{2}.
\end{equation}
The vortex radius can be roughly considered as
\begin{equation}
  r=2.1,
\end{equation}
with $\phi(2.1)\approx0.824$. The assignment here is for the convenience of later discussion. The exact value of vortex radius is determined by the behavior of Magnus force in Sec.\ref{sube}.

\subsection{vortices annihilation process}
\label{subv}

\begin{figure}
		\centering
		\includegraphics[scale=0.3]{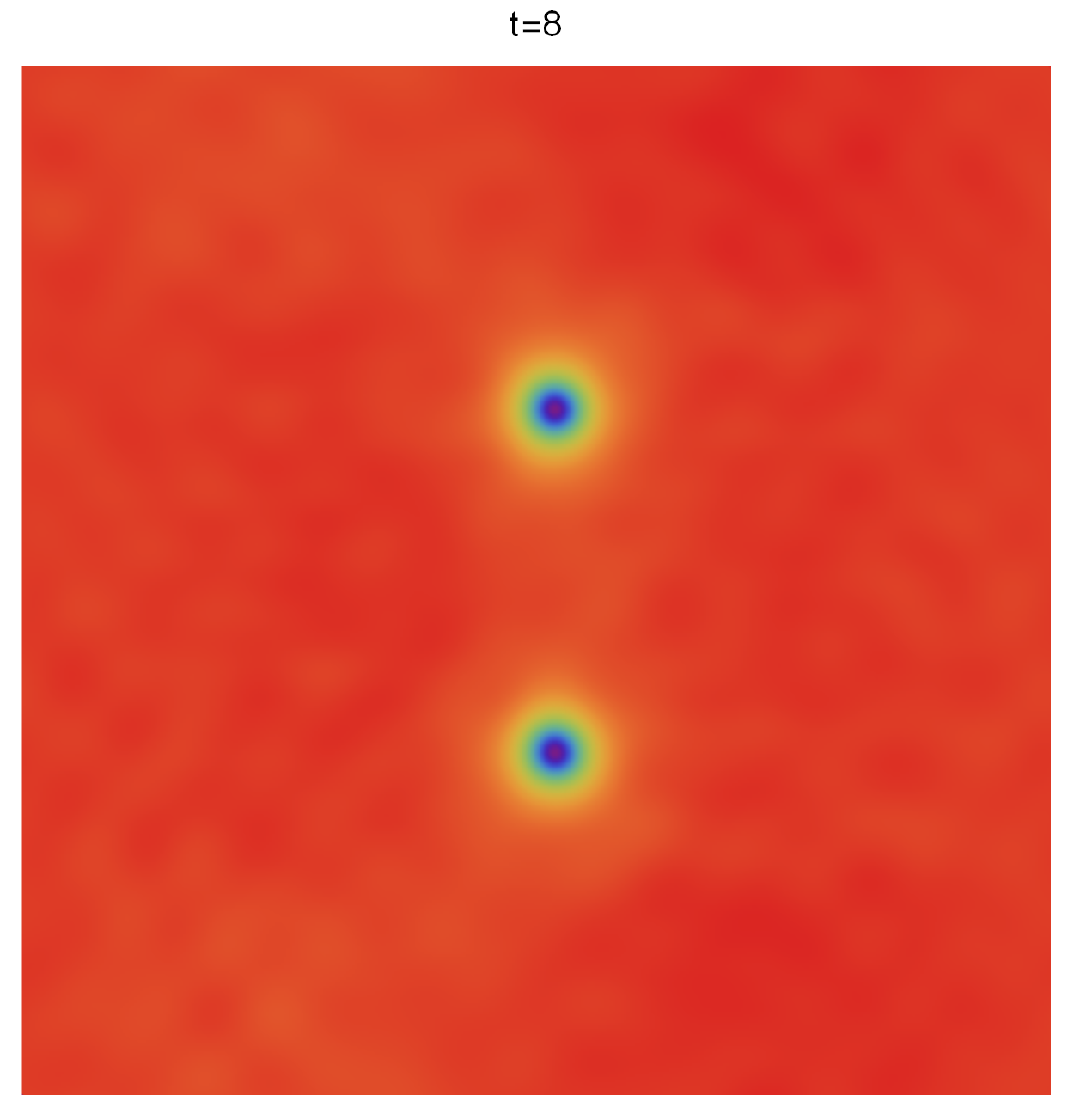}
        \includegraphics[scale=0.3]{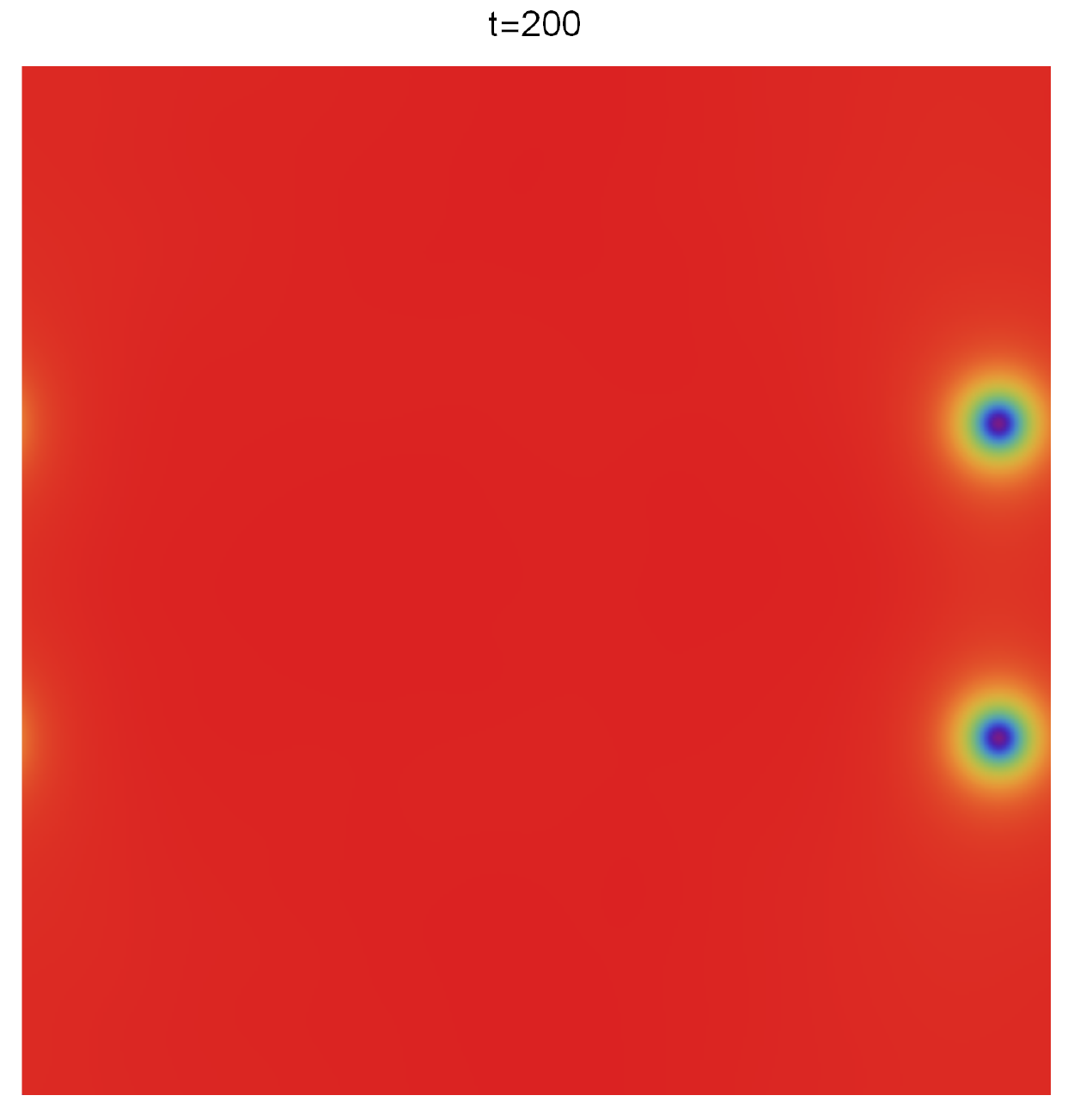}
        \includegraphics[scale=0.3]{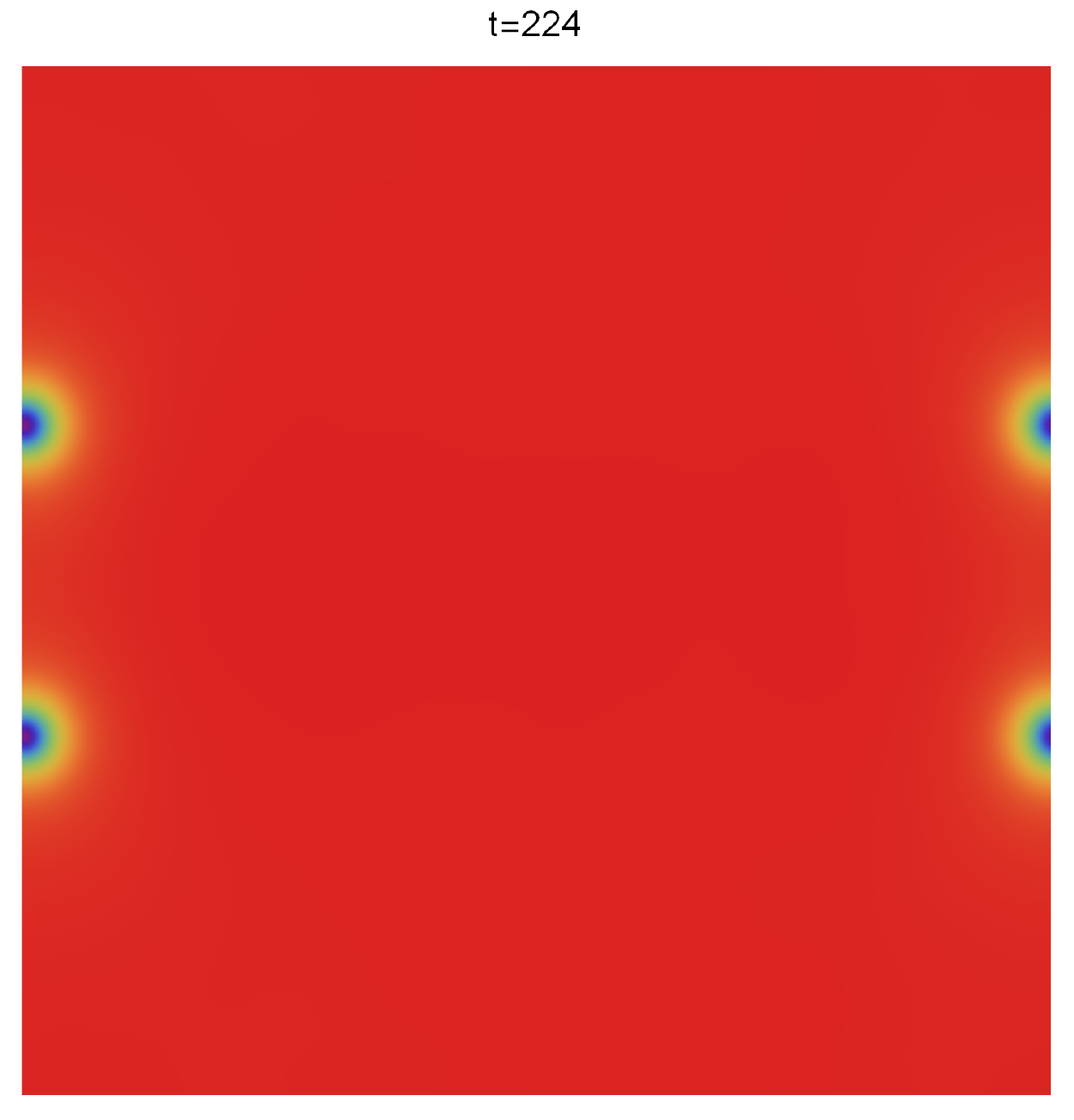}
        \includegraphics[scale=0.3]{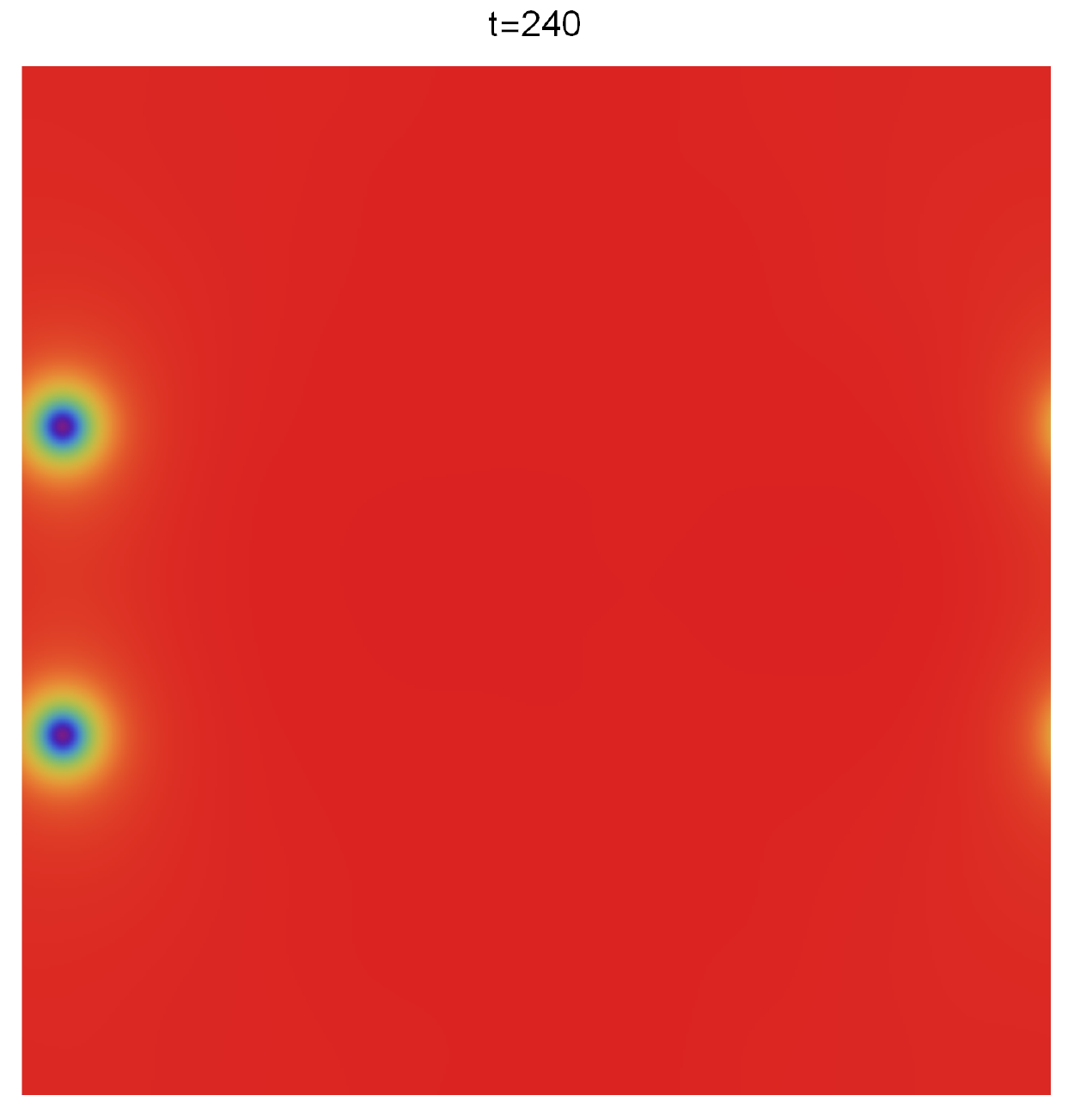}
        \includegraphics[scale=0.3]{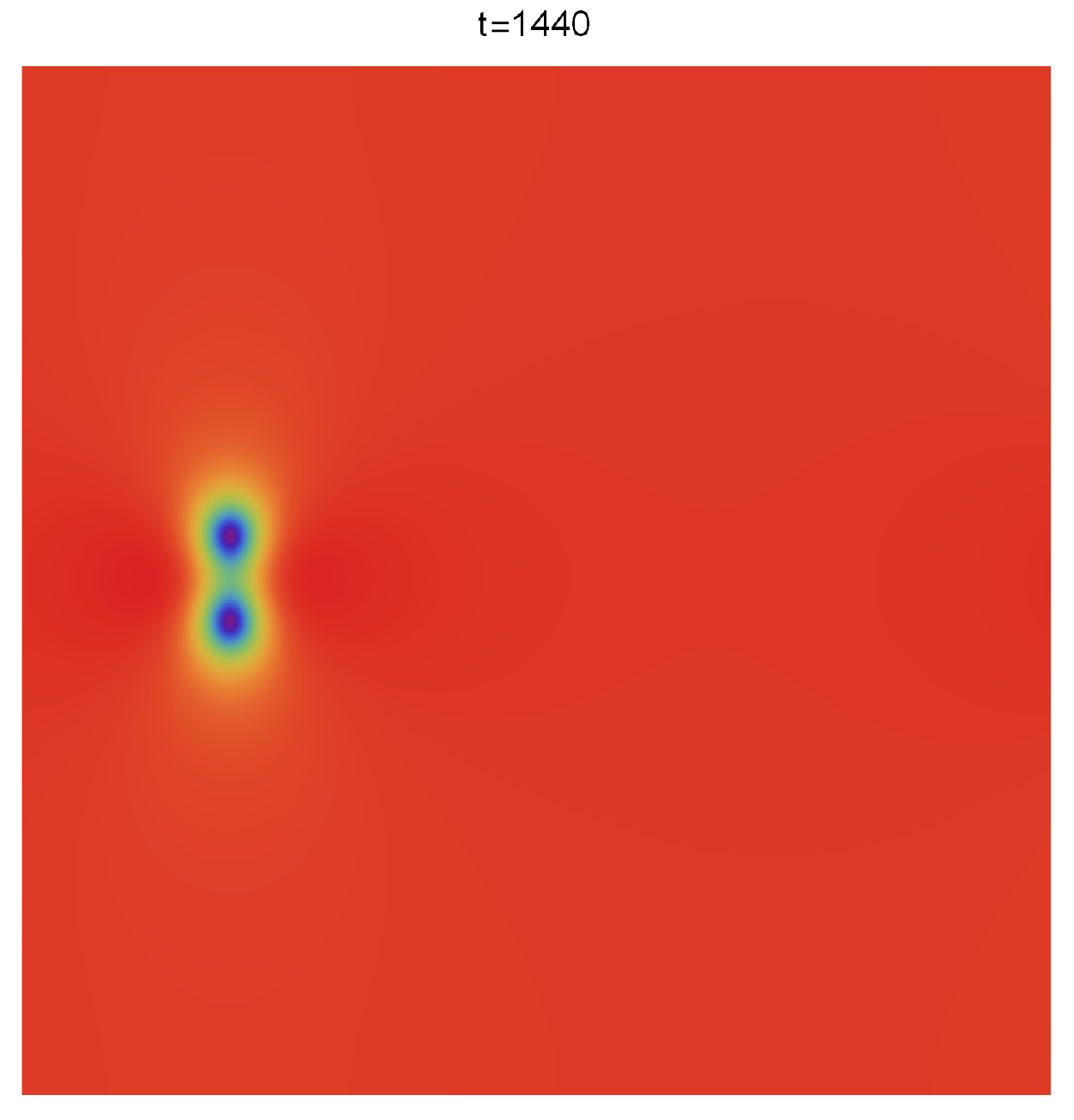}
        		\includegraphics[scale=0.3]{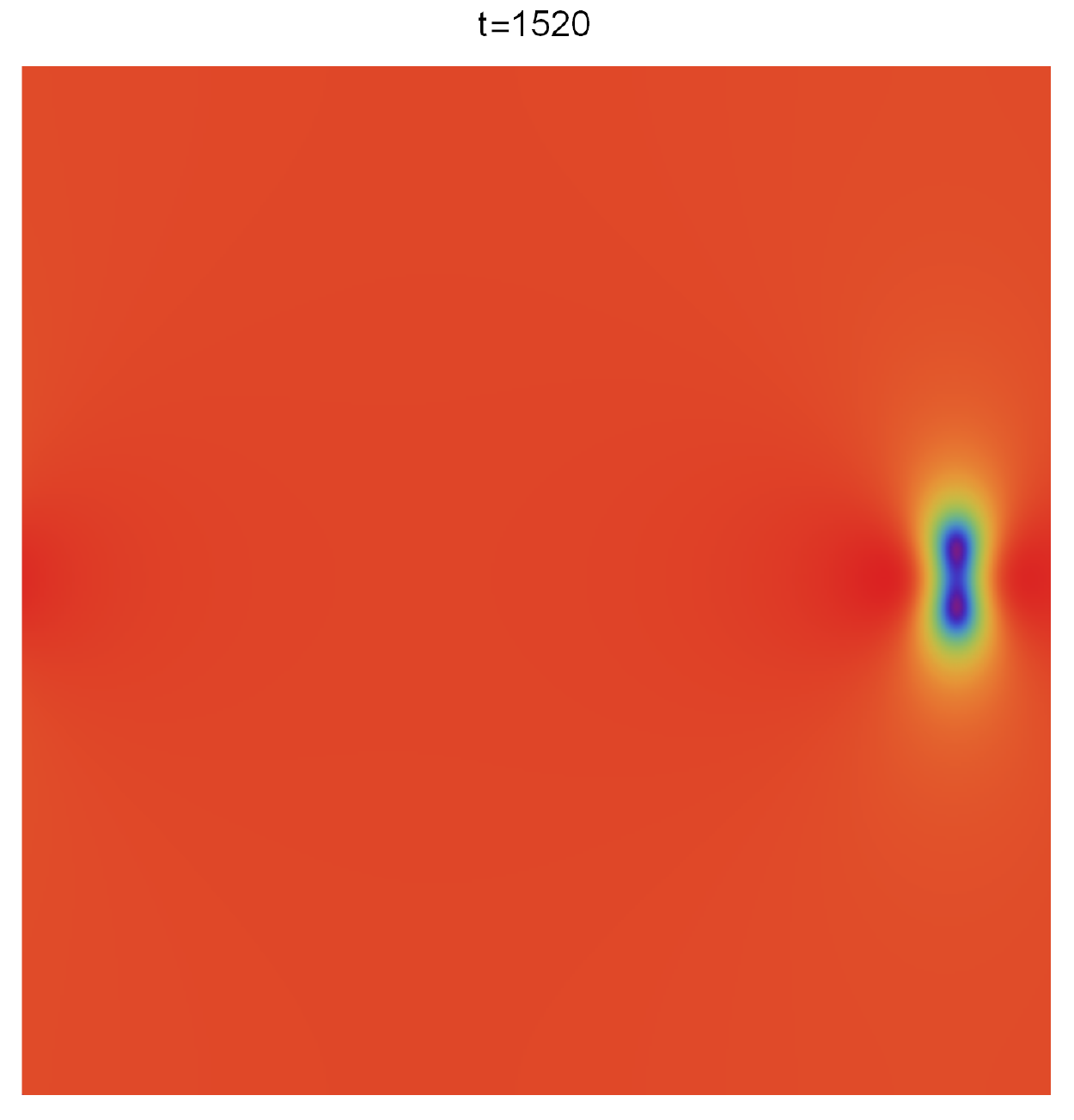}
        \includegraphics[scale=0.3]{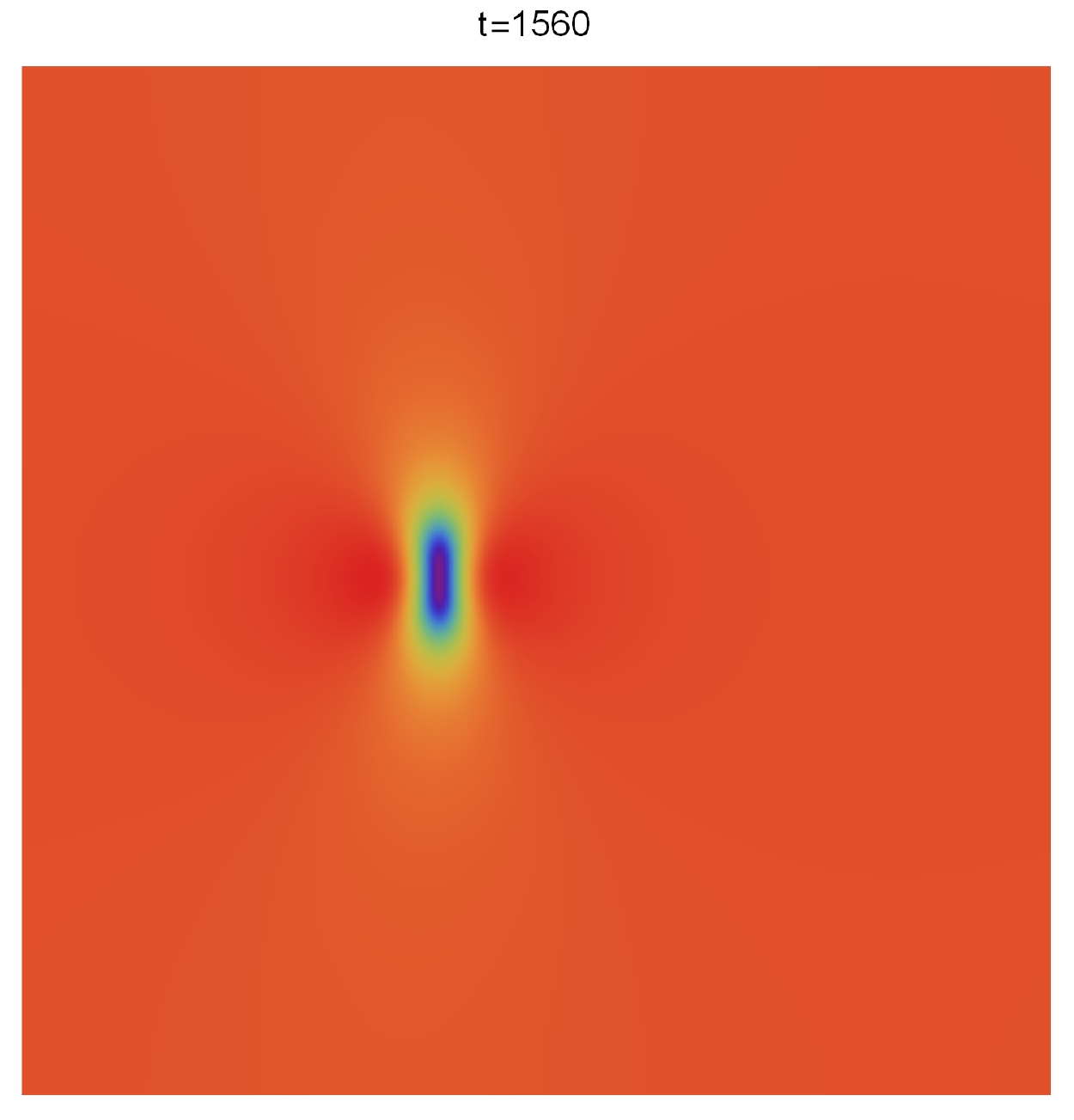}
        \includegraphics[scale=0.3]{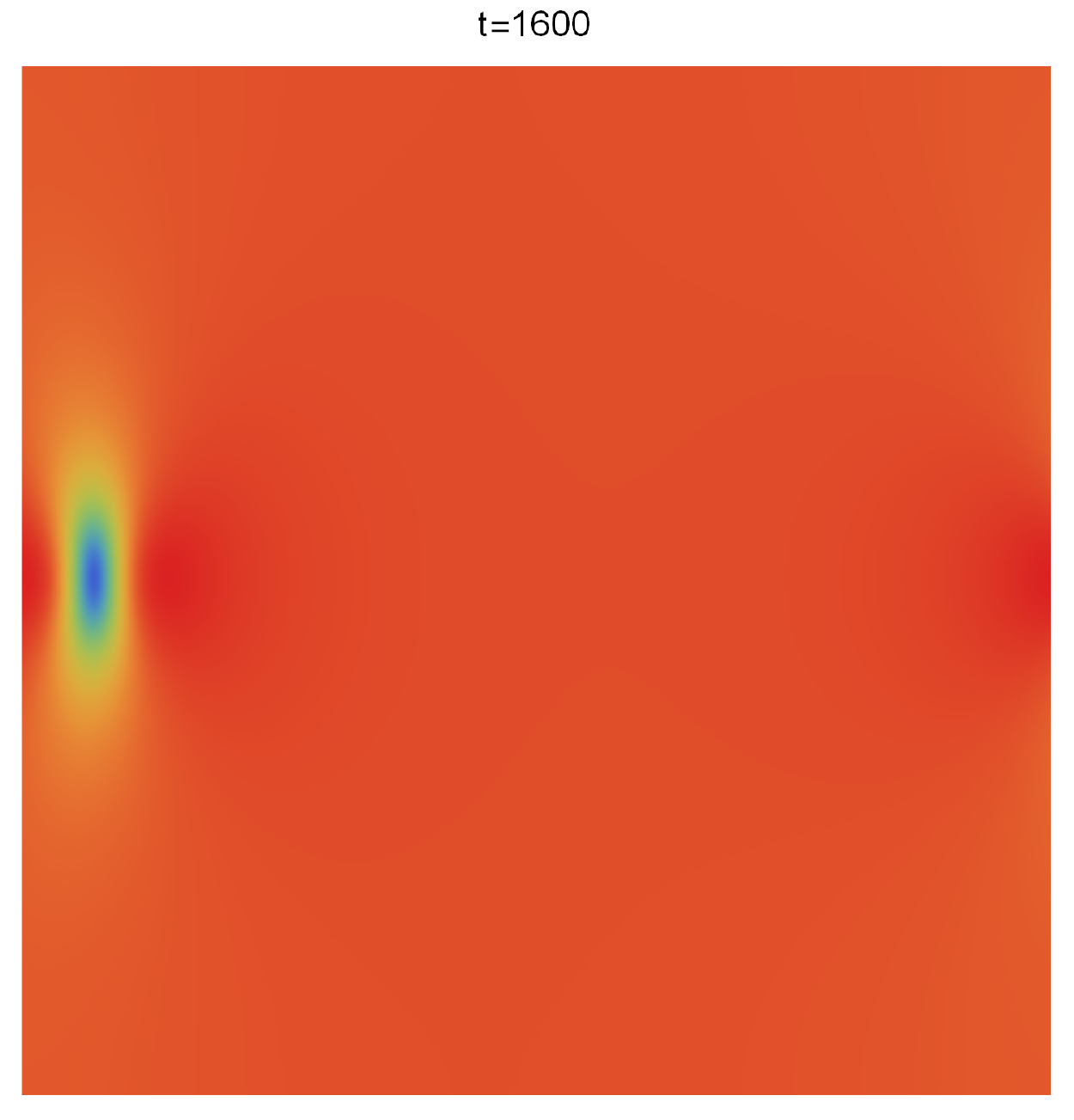}
        \includegraphics[scale=0.3]{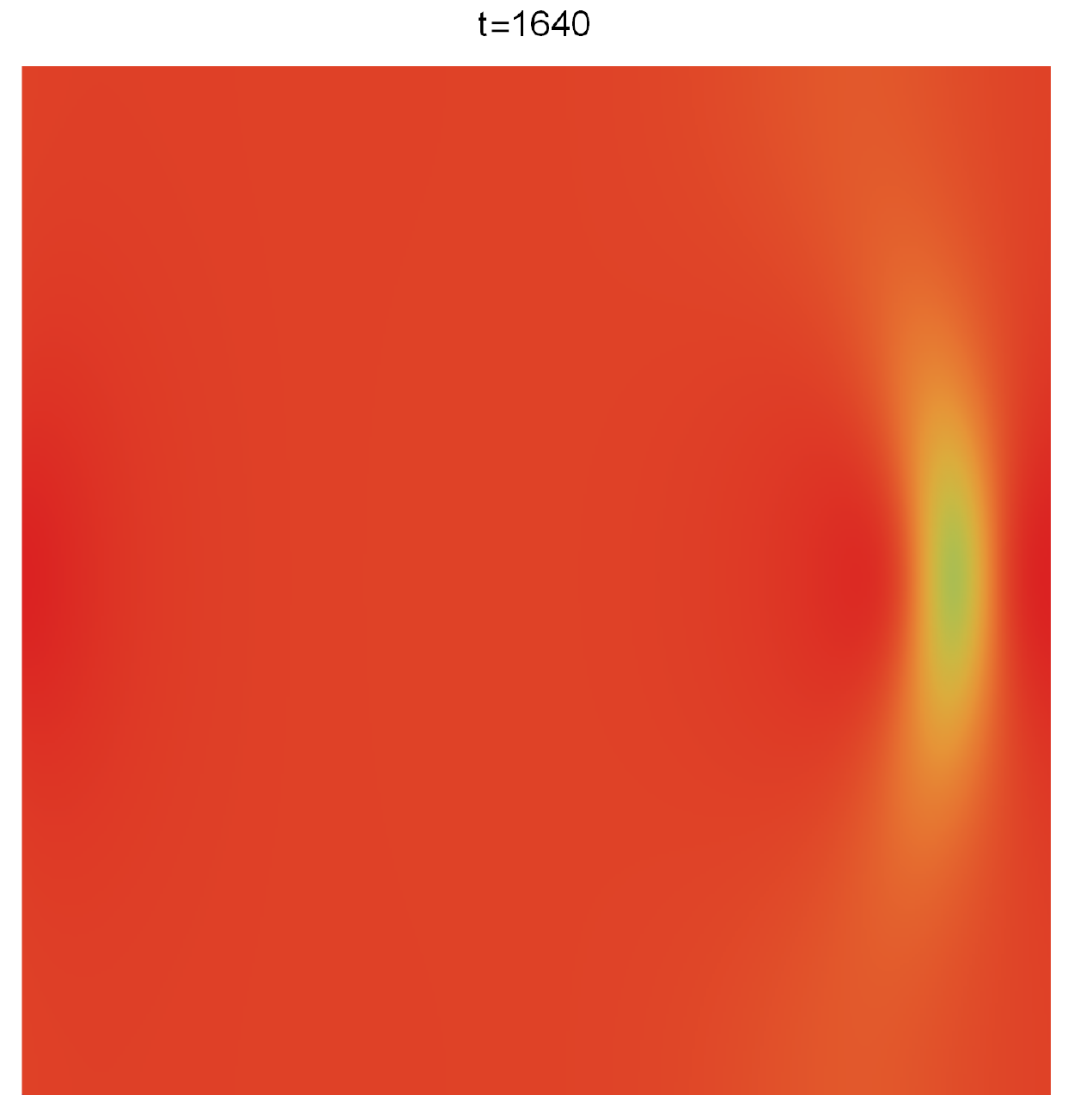}
        \includegraphics[scale=0.3]{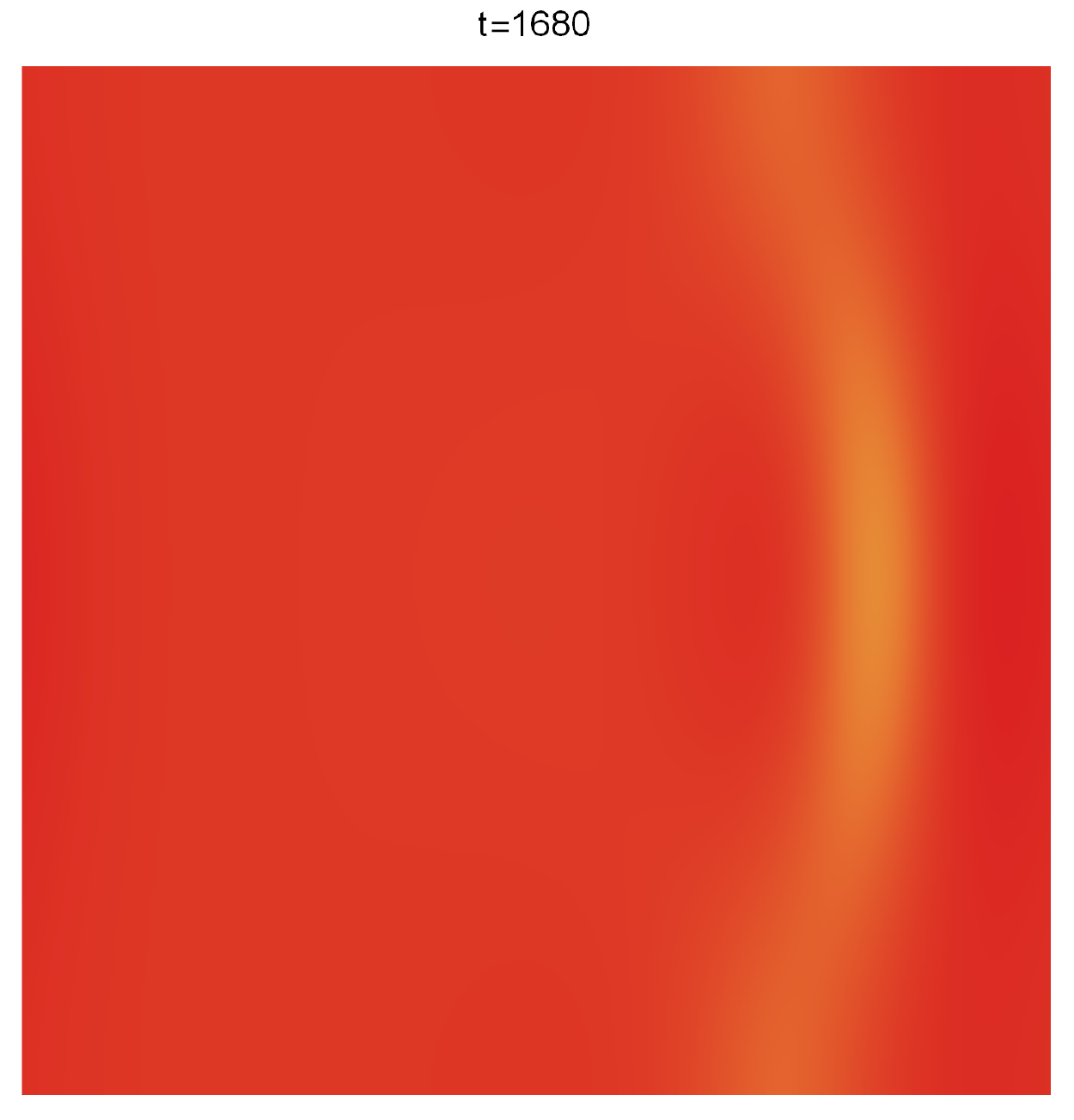}
        \includegraphics[scale=0.3]{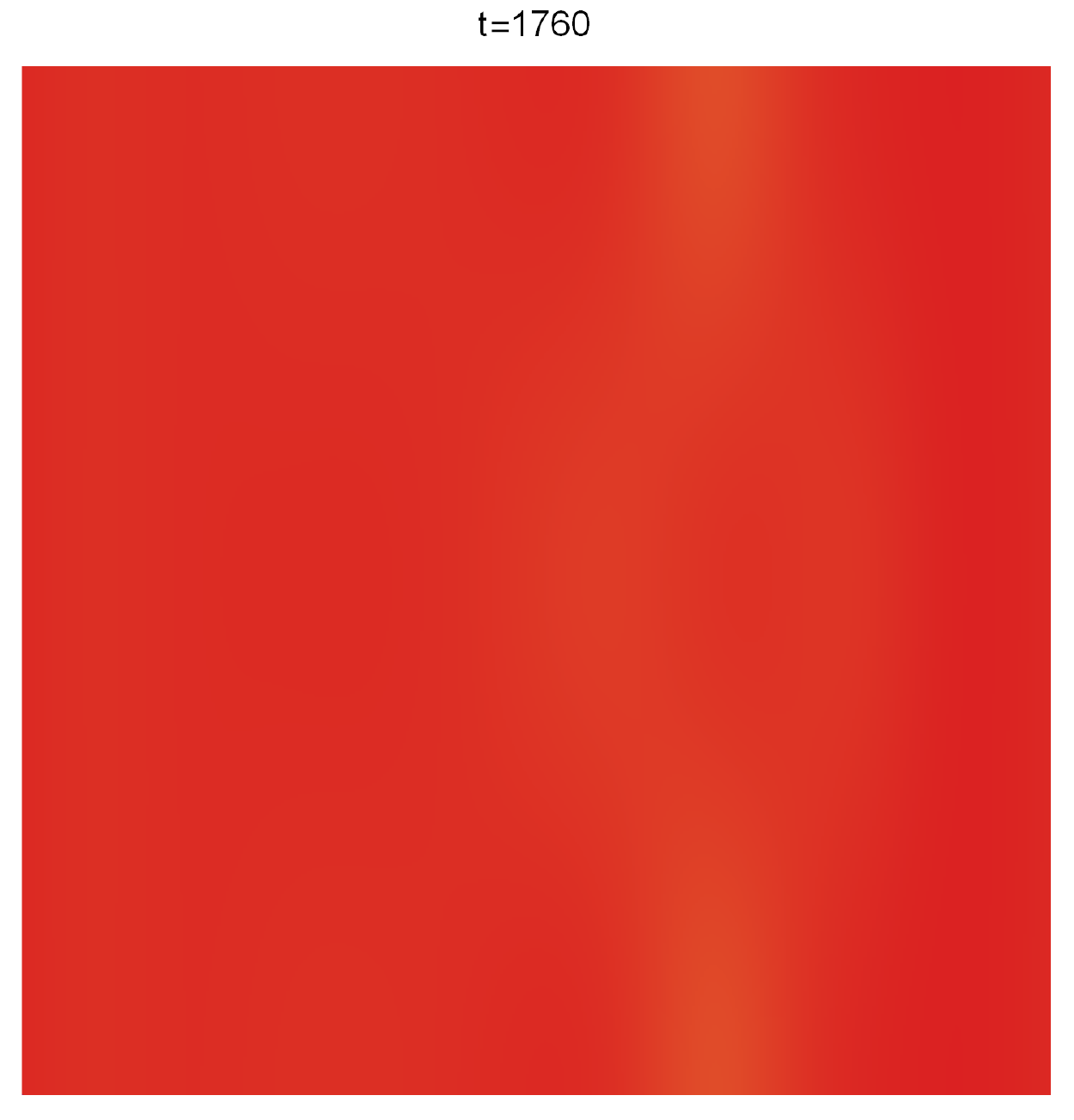}
        \includegraphics[scale=0.45]{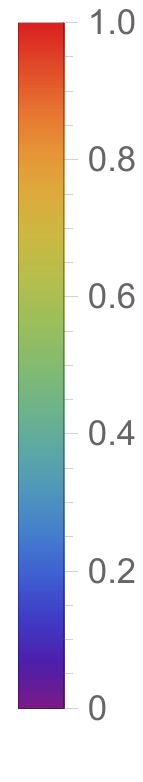}
	\caption{The vortices' annihilation process for $\eta=0.01$. The panels are superfluid configurations for different time . The dots are vortices, above is a positive vortex and below is a negative vortex. The positive and negative vortex move in the same positive x direction, and meantime they move towards each other until they annihilate into a soliton and then a crescent-shaped shock wave. Note the periodic boundary.}
	\label{vortime}
\end{figure}

We achieve long-time simulations of the vortices' annihilation process for different dissipative parameters ($\eta=0.01,0.02,0.05,0.1$). These simulations have very similar behavior except different time scales. As an example, $\eta=0.01$ case is shown in Fig.\ref{vortime}. The panels are superfluid configurations for different time $t=8, 200, 224, 240, 1440, 1520, 1560, 1600, 1640, 1680, 1760$. The dots are vortices, above is a positive vortex and below is a negative vortex. As one can see, the positive and negative vortex move in the same positive x direction, and meantime they move towards each other until they annihilate into a soliton and then a crescent-shaped shock wave.

\begin{figure}
		\centering
		\includegraphics[scale=0.45]{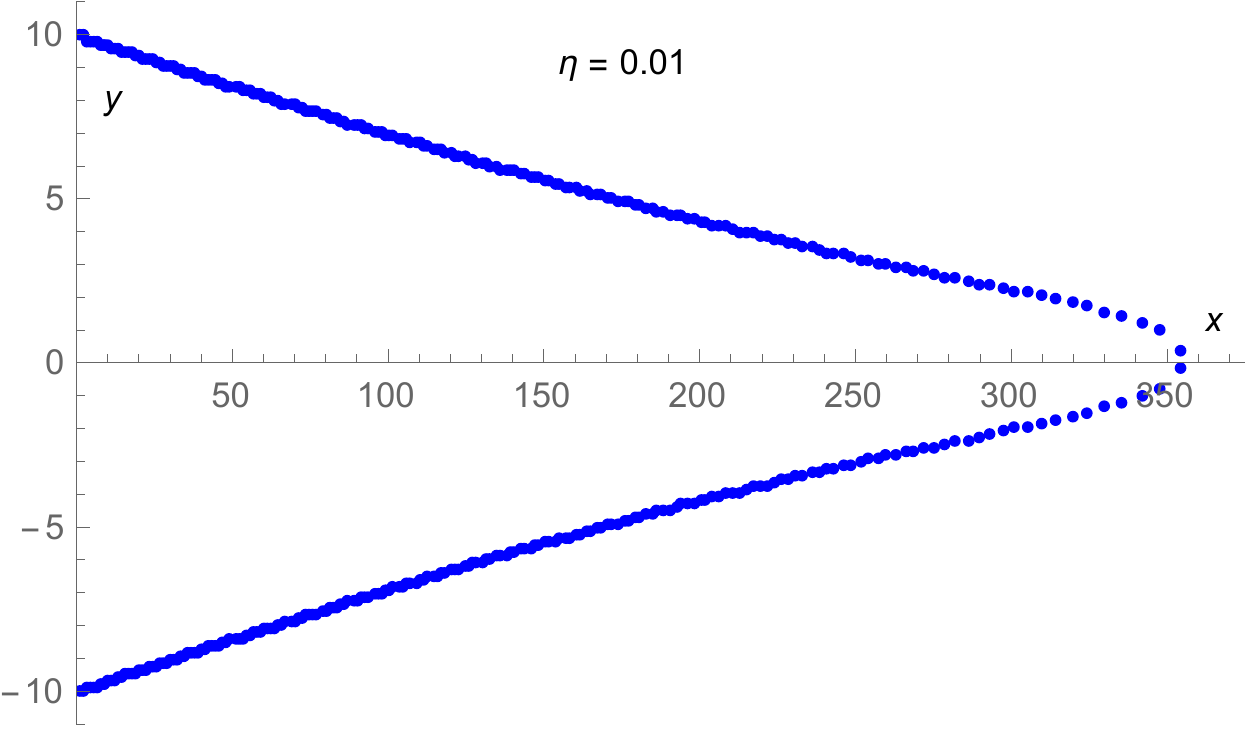}
        \includegraphics[scale=0.45]{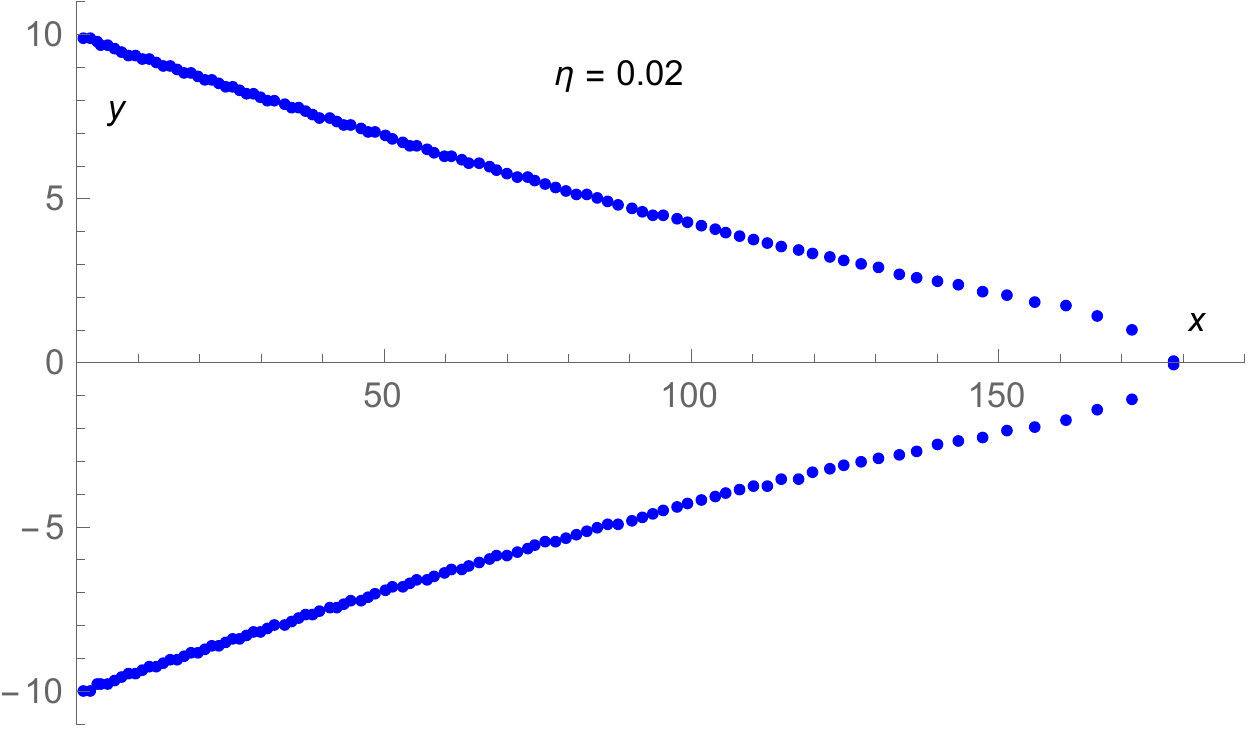}
        \includegraphics[scale=0.45]{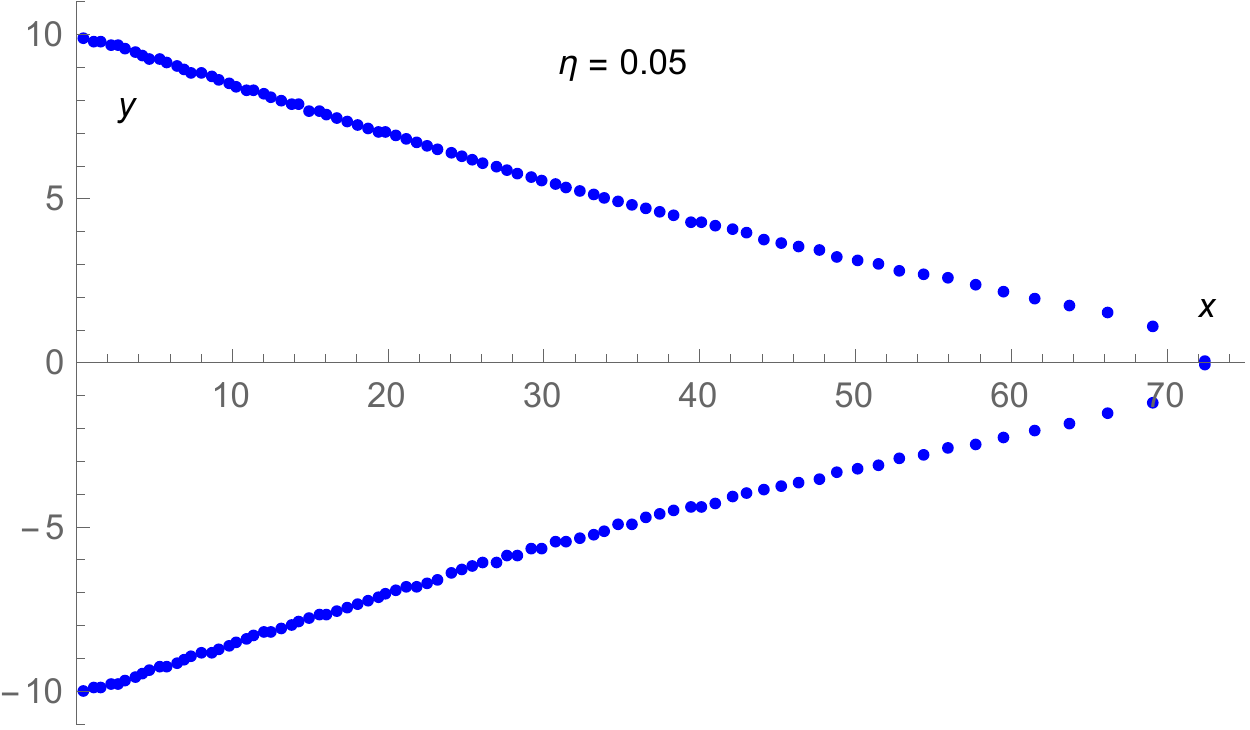}
        \includegraphics[scale=0.45]{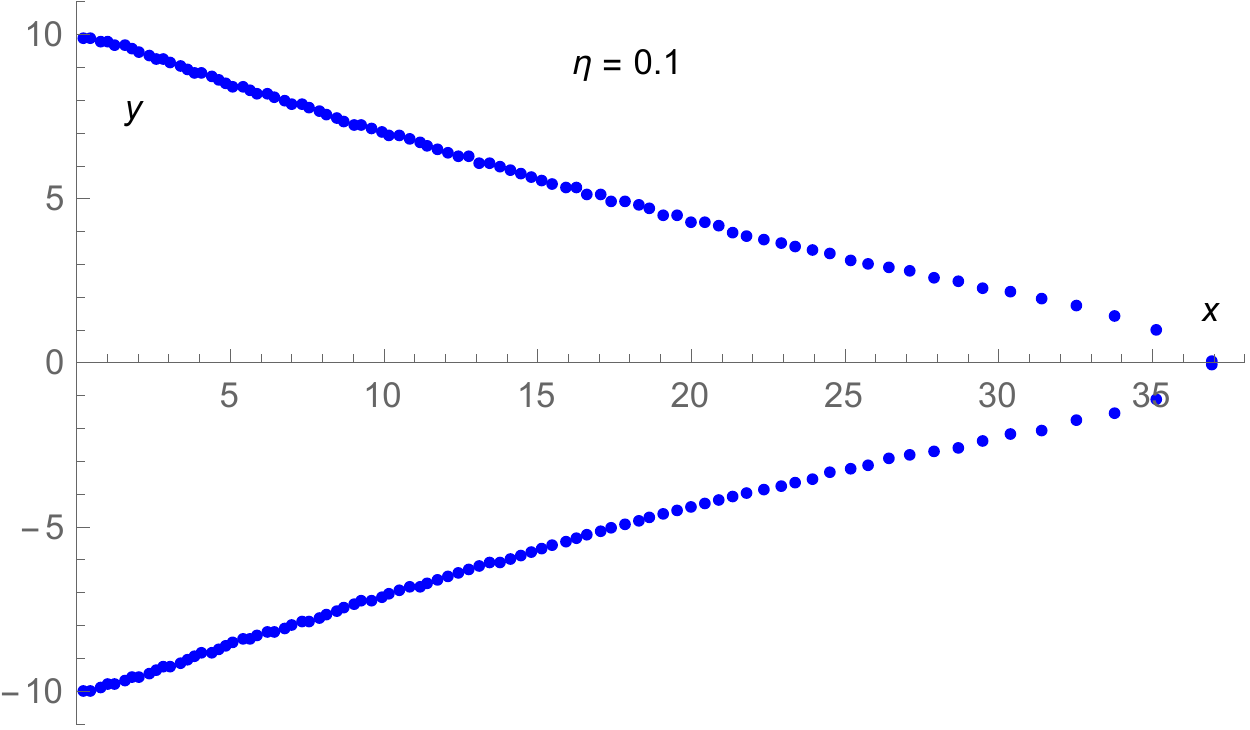}
	\caption{Trajectories of vortices in x-y plane for different dissipative parameters. The vortices move a longer distance in x axis before annihilating into a soliton for a smaller dissipative parameter. The motion of positive and negative vortices have very good symmetry about time and x axis.}
	\label{vorxy}
\end{figure}

During the simulation, we recorded the positions of the vortices every $\Delta t$.  Fig.\ref{vorxy} shows the trajectories of vortices corresponding to different dissipative parameters. For the case of small dissipative parameter, the vortices move a long distance in x axis before annihilating into a soliton. One can also find that for each simulation, the motion of  positive and negative vortices have very good symmetry about time and x axis. Therefore, we only need to consider the motion of one of the vortices.

\begin{figure}
		\centering
		\includegraphics[scale=0.45]{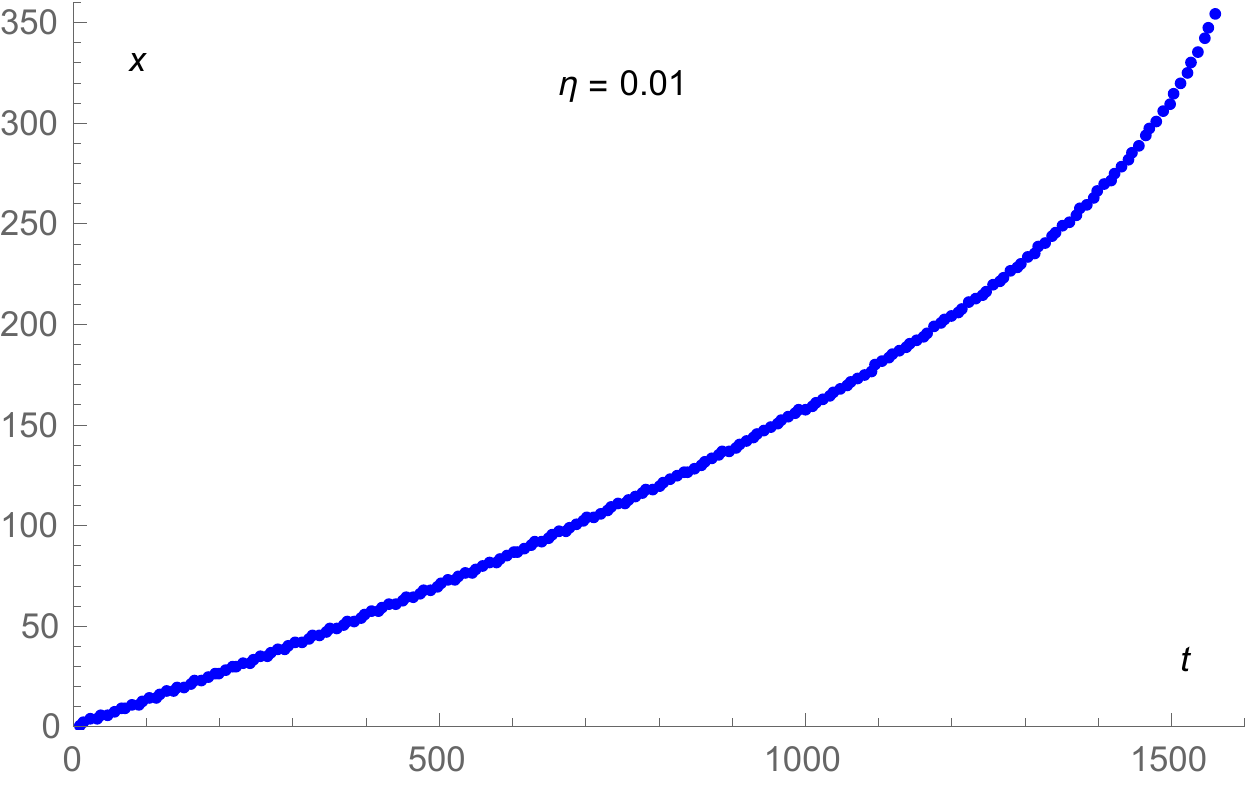}
        \includegraphics[scale=0.45]{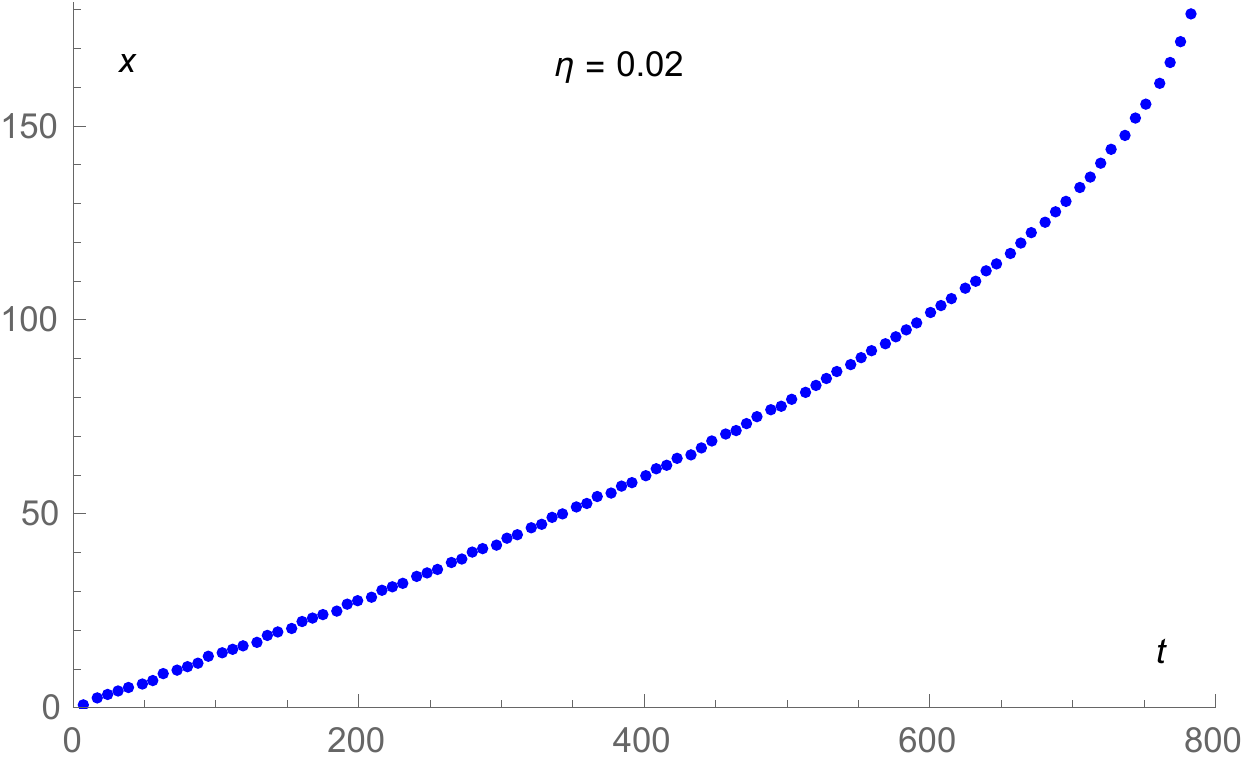}
        \includegraphics[scale=0.45]{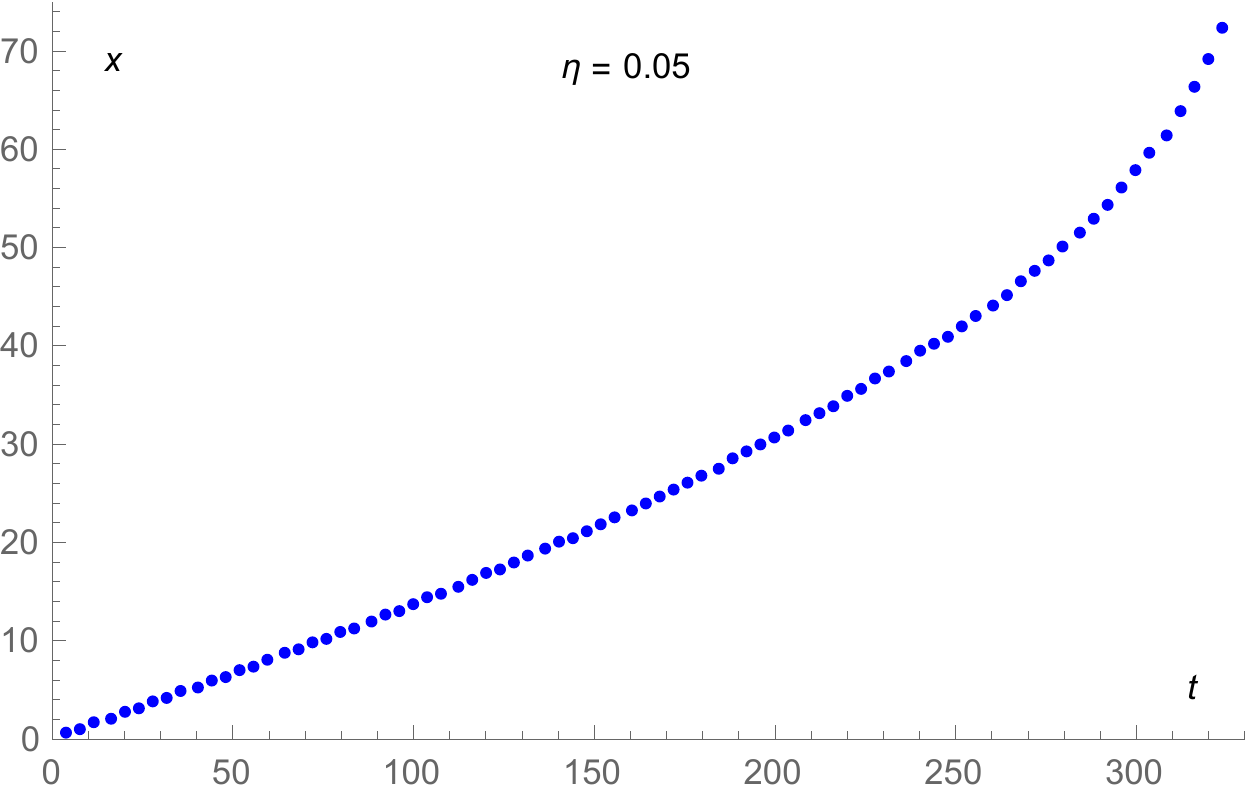}
        \includegraphics[scale=0.45]{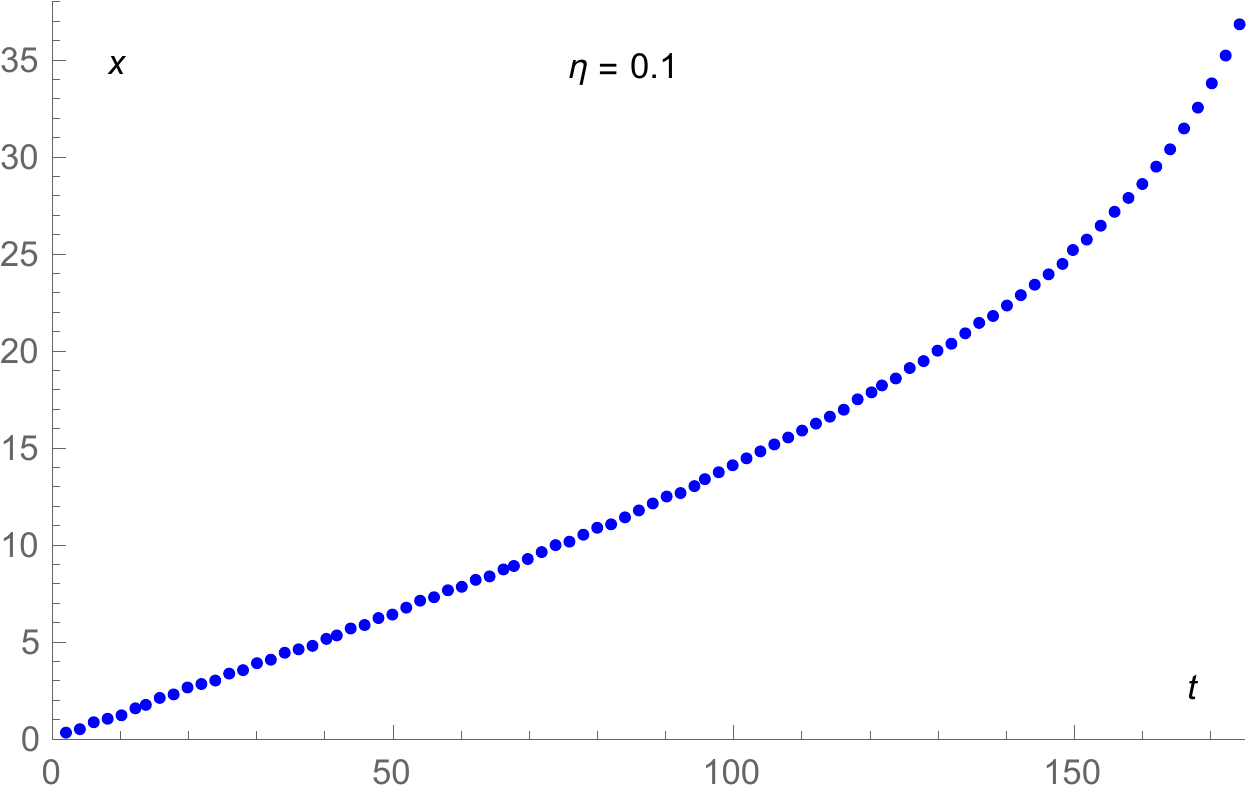}
	\caption{ x vs. t behavior of vortices for different dissipative parameters. Despite the different length scale in x axis and time scale, the four graphes are similar. }
	\label{vorxt}
\end{figure}

Fig.\ref{vorxt} shows the x vs. t behavior of vortices for different dissipative parameters. One can see that the vortices accelerate in x direction as their velocities increase with time. The acceleration is slower for smaller dissipative parameter. Despite the different length scale in x axis and time scale, the four graphes are similar.

\begin{figure}
		\centering
		\includegraphics[scale=0.45]{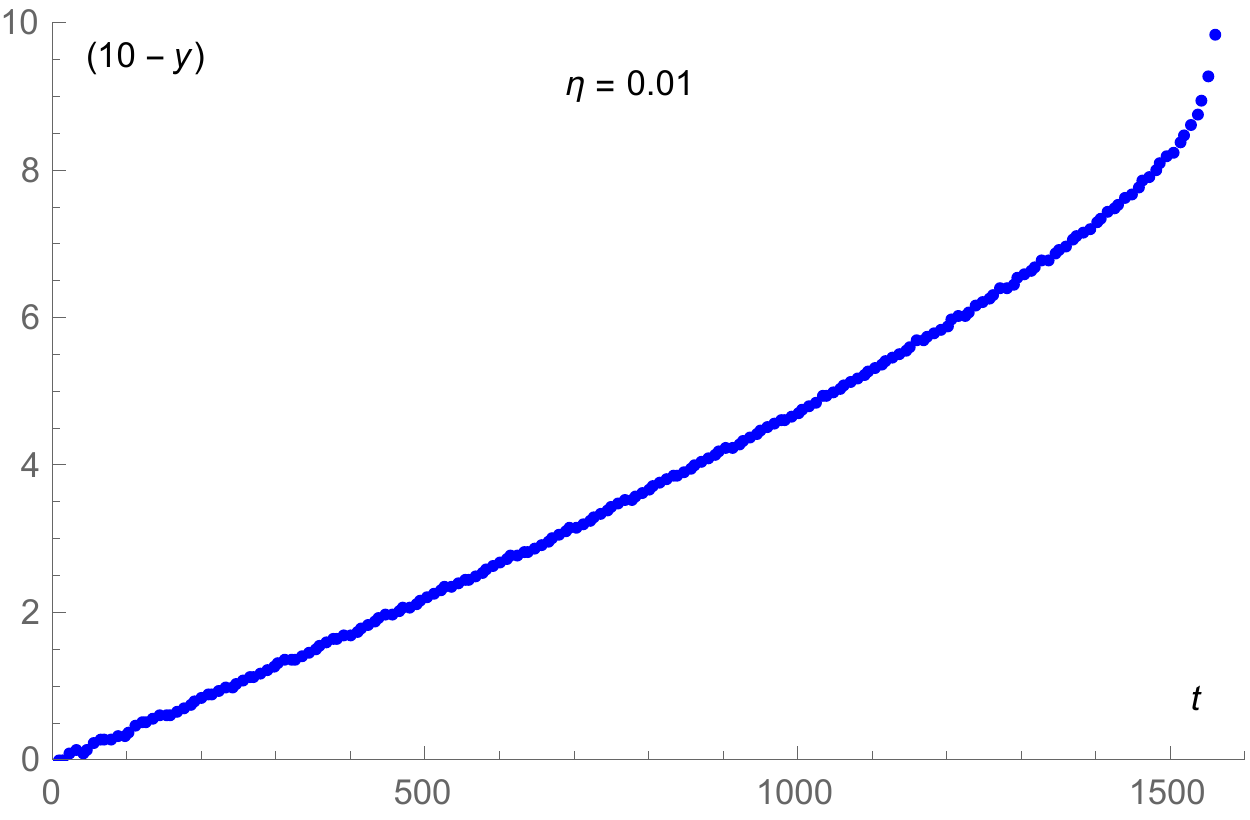}
        \includegraphics[scale=0.45]{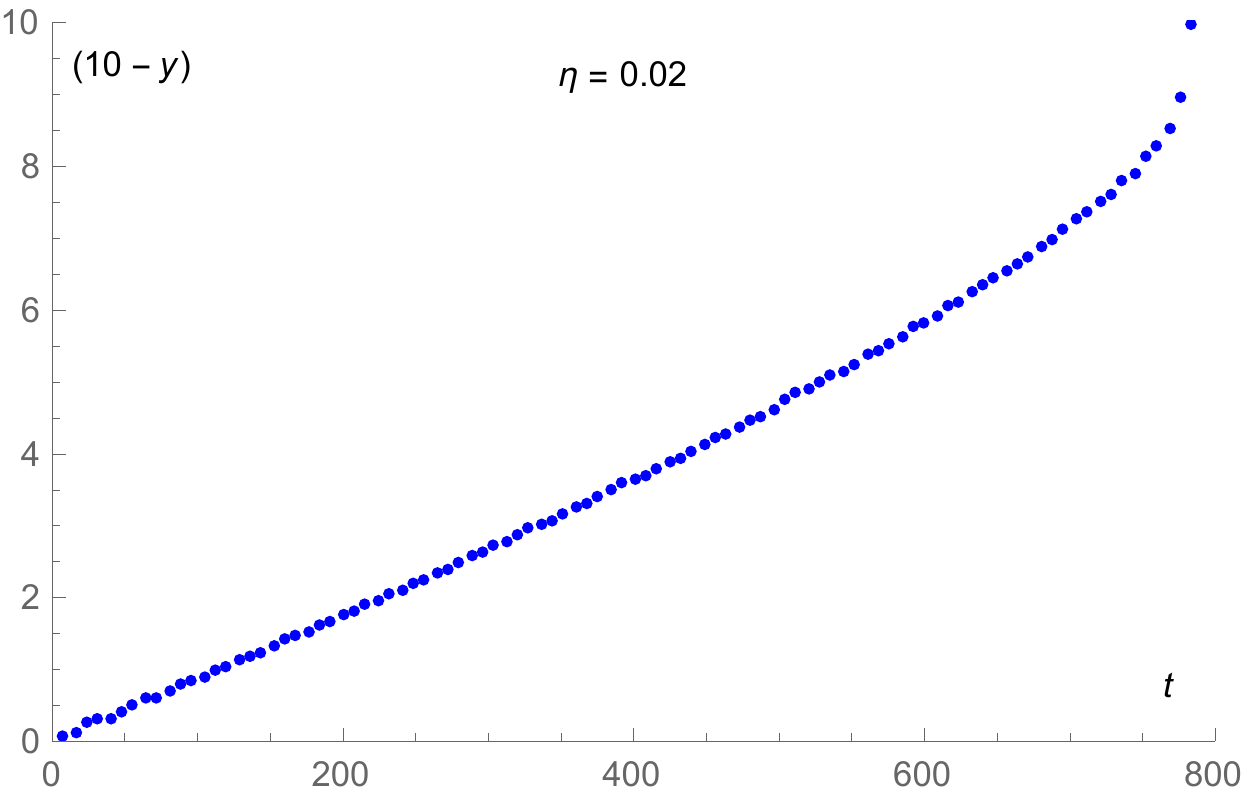}
        \includegraphics[scale=0.45]{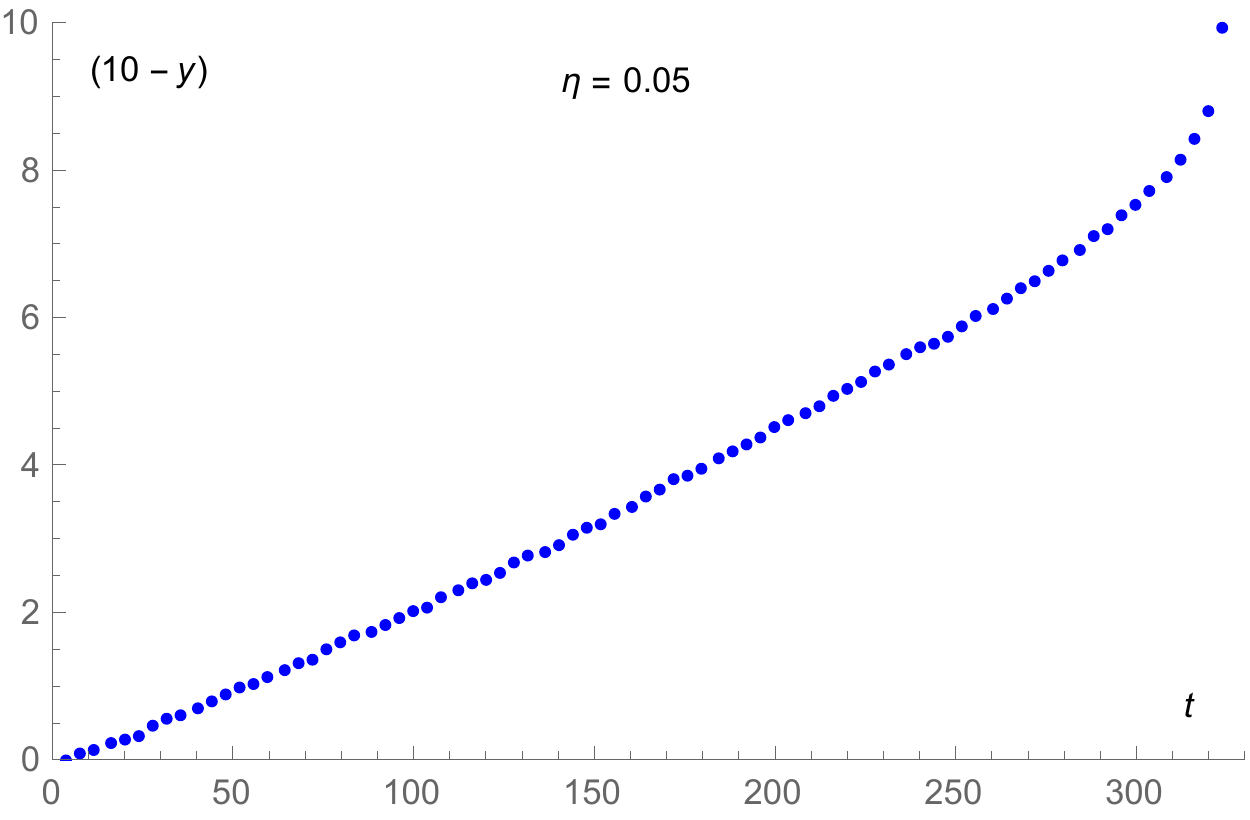}
        \includegraphics[scale=0.45]{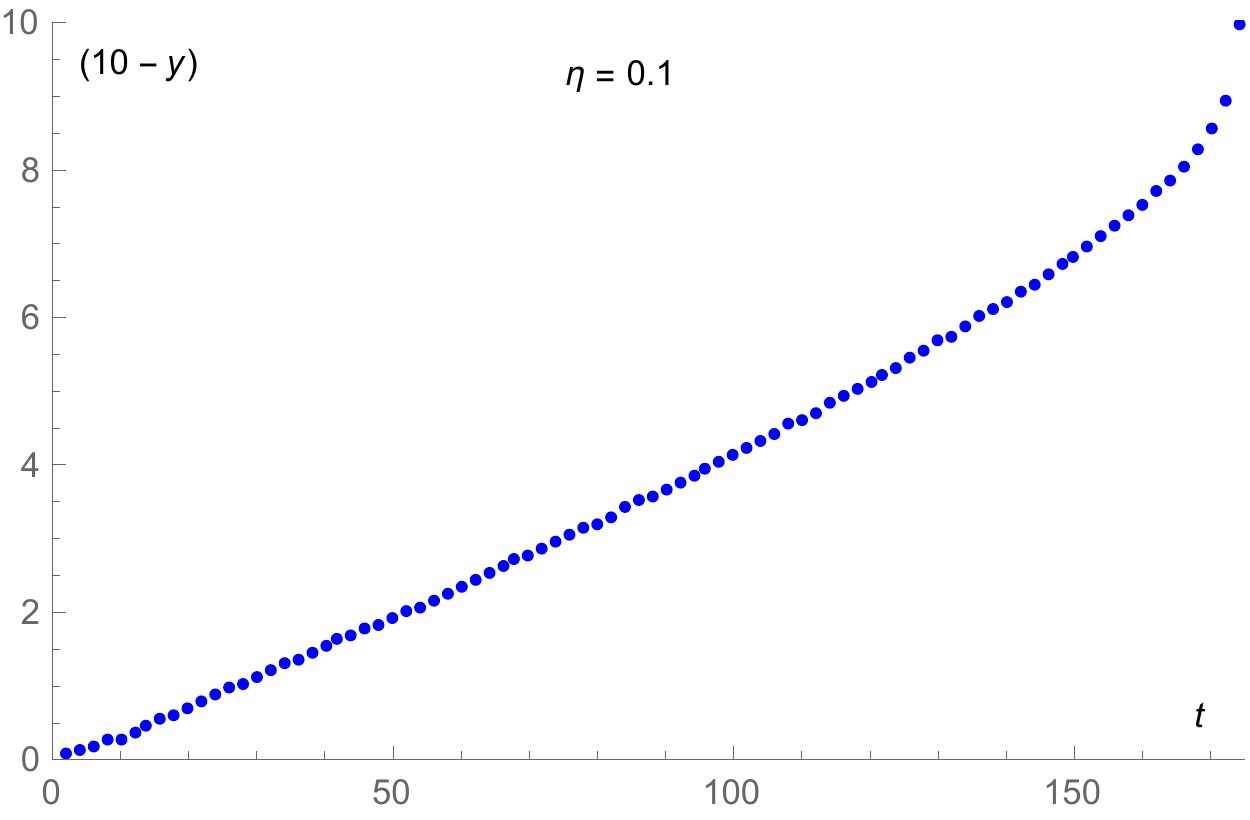}
	\caption{(10-y) vs. t behavior of positive vortices for different dissipative parameters. Despite the different time scale, the four graphes are similar. }
	\label{voryt}
\end{figure}

Fig.\ref{voryt} shows (10-y) vs. t behavior of positive vortices for different dissipative parameters. One can see that the vortices also accelerate in y direction as their velocities increase with time. The acceleration is slower for smaller dissipative parameter.  Despite the different time scale, the four graphes are similar.

Then how to understand the vortices' annihilation process, e.g. (1)why the vortices move towards positive x direction, (2)why they accelerate in both x and y directions, (3)why the accelerations are different for different dissipative parameters? As a poor fitting function will make us get inaccurate vortex velocity and acceleration and we haven't found the perfect fitting functions for the above graphes, we will only analyze the above phenomena qualitatively rather than quantitatively.

(1) The positive vortex generates counterclockwise rotating velocity fields around it and the negative vortex generates clockwise rotating velocity fields around it. Note that above is a positive vortex and below is a negative vortex. So the positive vortex is in positive x direction velocity field of the below negative vortex and the negative vortex is in positive x direction velocity field of the above positive vortex. In a result, both the vortices will move in positive x direction. If the positions of the positive and negative vortices are swapped, they will move in the negative x direction.

(2) When vortex drifts, there will be a Magnus force exerts on it\cite{ueda2010}. The force is in the direction perpendicular to both the vorticity and the drift velocity of the vortex. In our case, the vortices drift in positive x direction. The Magnus force which exerts on the positive vortex is in negative y direction and the Magnus force which exerts on the negative vortex is in positive y direction. So the vortices will accelerate toward each other in y direction. This makes the separation distance between the vortices smaller and the drift velocities bigger which in return make the Magnus forces bigger. Note that the Magnus force is defined when the separation distance of vortices is large. While when the separation distance of vortices is small, there still is a force as the vortices accelerate. Thus we still call it the Magnus force.

(3) The curves in Fig.\ref{vorxy}, Fig.\ref{vorxt} and Fig.\ref{voryt} are similar for different dissipative parameters. While their time scales $\tau$ are different. It seems that $\tau \eta\approx 15.81, 15.88, 16.08, 17.5$ are constant, or $\tau \propto 1/\eta$.  As we already know in the above paragraph that the accelerations of vortices are caused by Magnus forces, then different accelerations for different dissipative parameters seem to denote different Magnus forces. While it is not true as the Magnus force is configuration dependent which can be saw in next subsection. The acceleration in x direction increases with dissipative parameter should be attributed to a decrease of ``vortex inertia mass".

\begin{figure}
		\centering
		\includegraphics[scale=0.45]{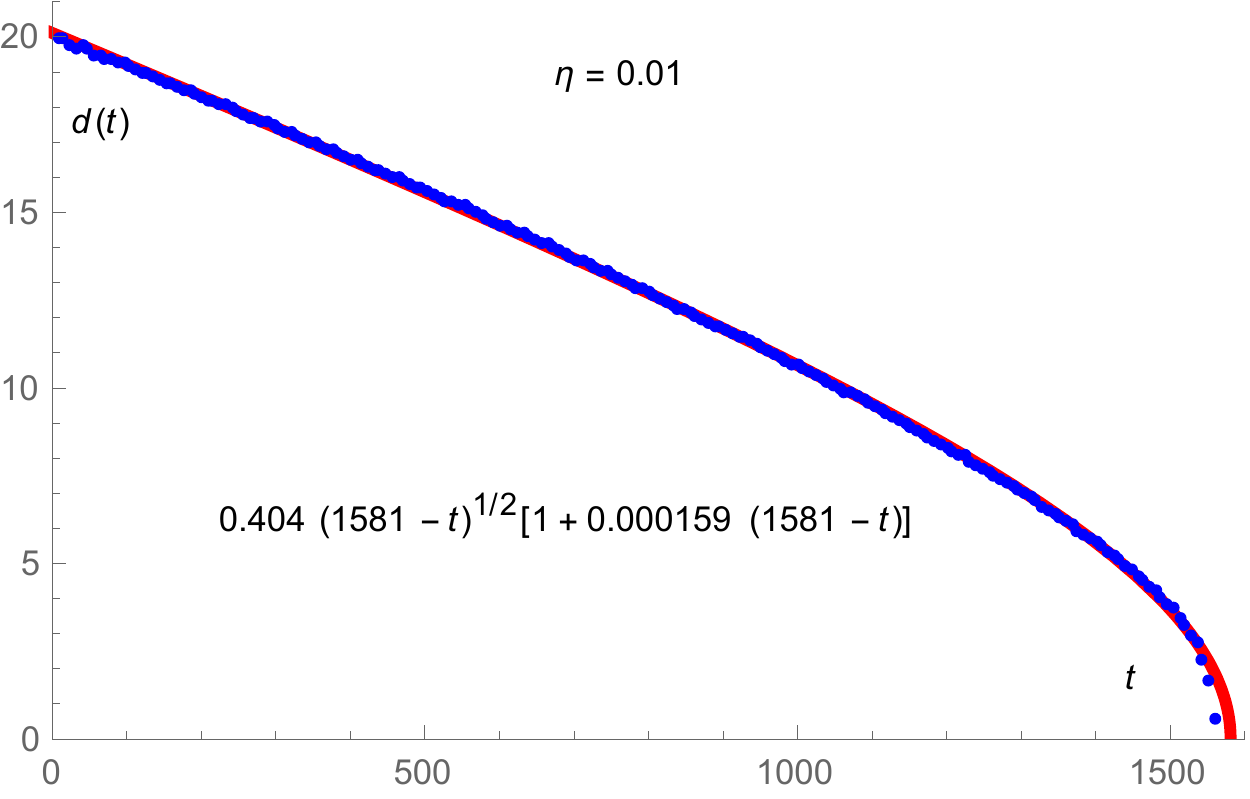}
        \includegraphics[scale=0.45]{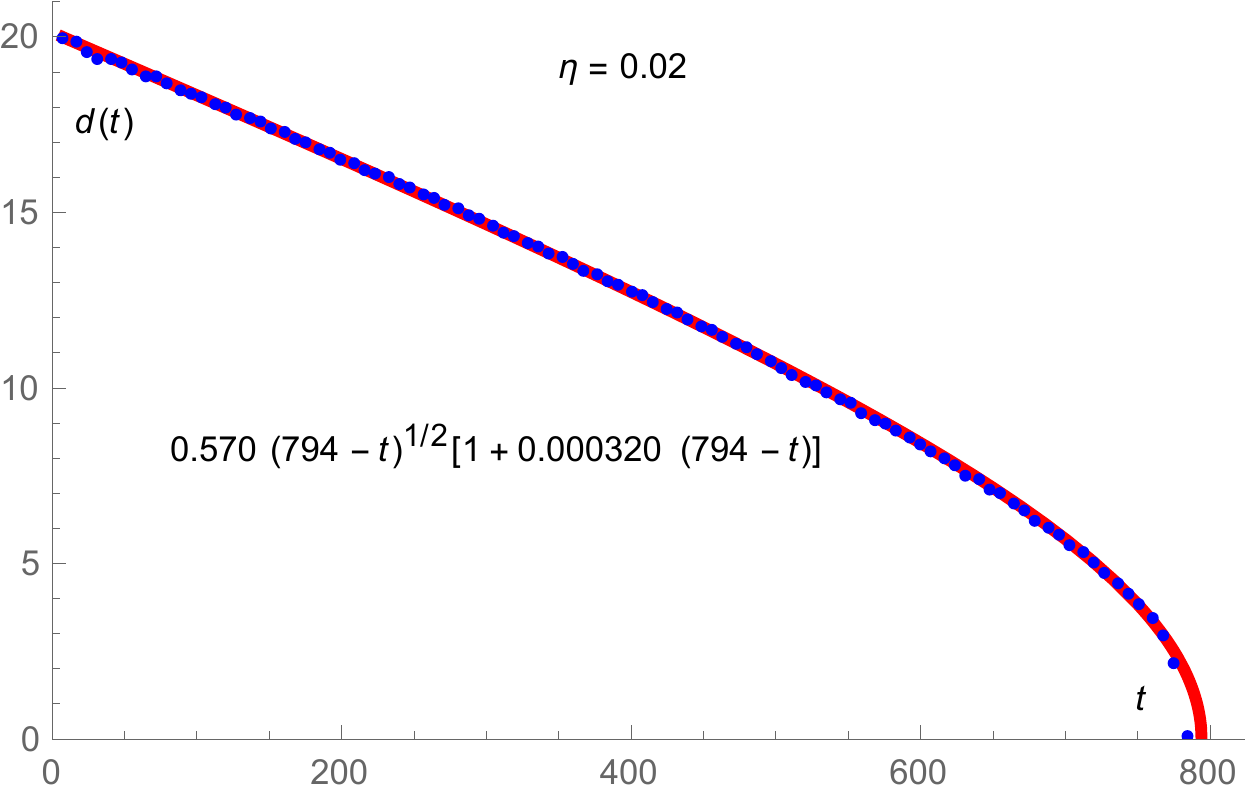}
        \includegraphics[scale=0.45]{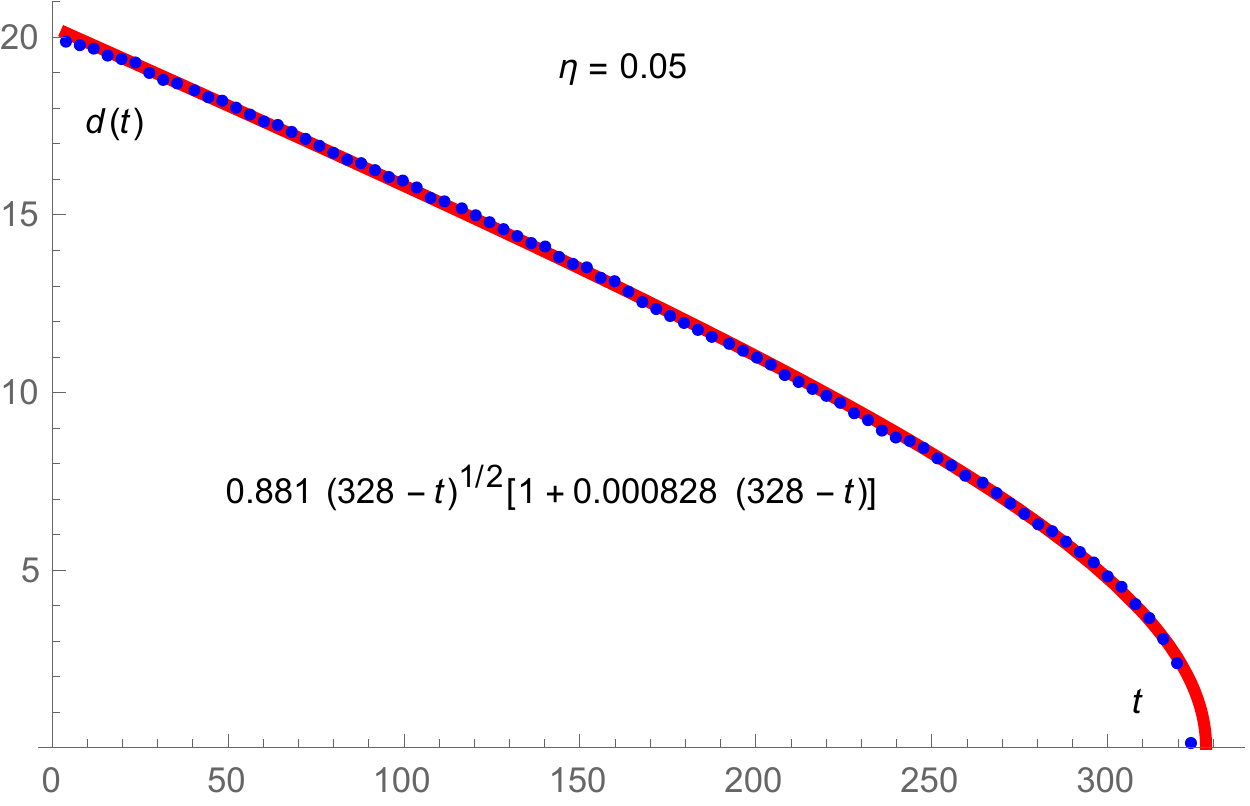}
        \includegraphics[scale=0.45]{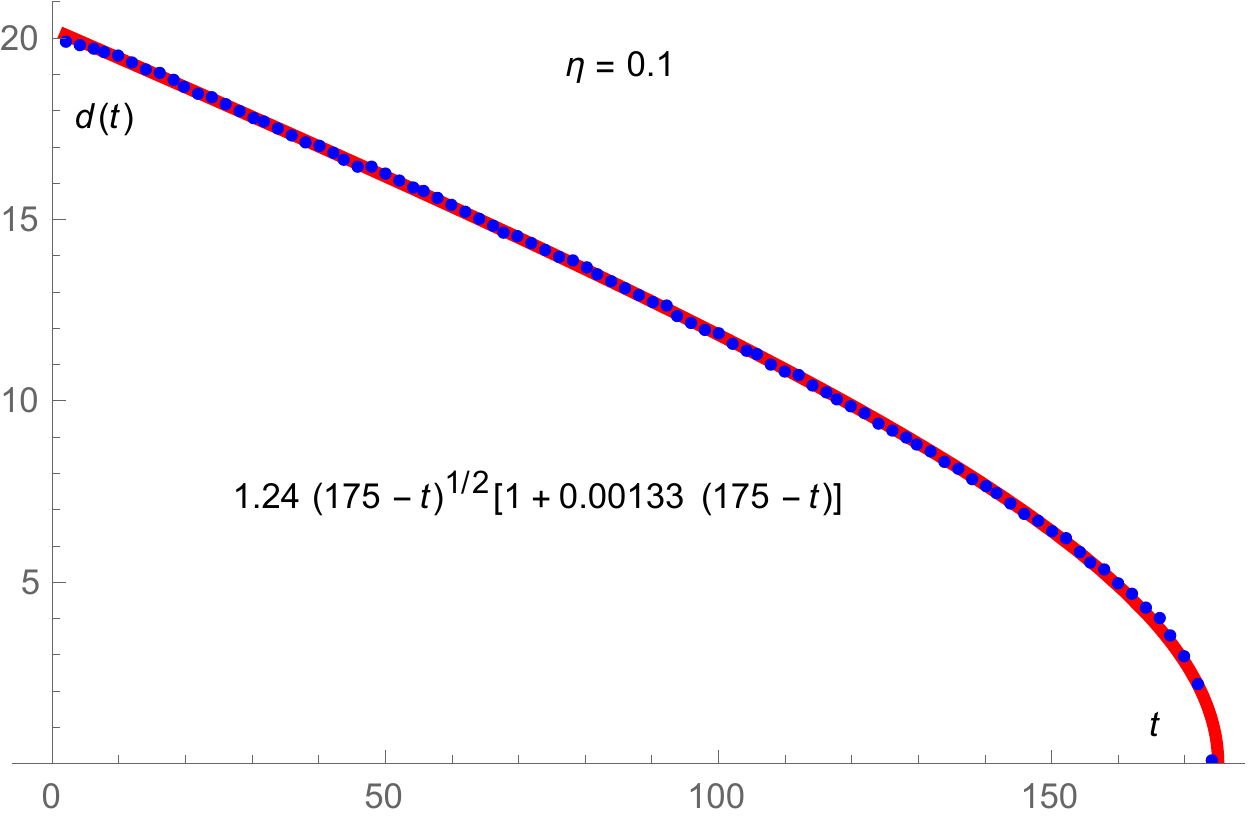}
	\caption{The blue dots are $d(t)$ vs. t, where $d(t)$ is the separation distance between the vortices at different times. The red curves are fitting functions $d(t)=B_{1}(t_{0}-t)^{1/2}[1+B_{2}(t_{0}-t)]$. }
	\label{dist}
\end{figure}

Fig.\ref{dist} shows $d(t)$ vs. t, where $d(t)$ is the separation distance between the vortices at different times. We find that function $d(t)$ are roughly fitted by the formula
\begin{equation}
  d(t)=A_{1}(t_{0}-t)^{1/2},
\end{equation}
where $t_{0}$ is the reconnection time. While they are approximately fitted by
\begin{equation}
  d(t)=B_{1}(t_{0}-t)^{1/2}[1+B_{2}(t_{0}-t)].
\end{equation}
Specially, when the separation distances are large, e.g. $d(t)>4.2$ ( the 4.2 can be considered as the length scale of the vortex which can be confirmed in Fig.\ref{svortex} ), their behavior are well fitted by the above formula. Note that $B_{2}$ are very small constants, the scaling law $1/2$ of the separation distance between vortices seems to be universal for two dimensional and three dimensional cases\cite{1994prlwaele,2008PNASbewley,2008pdnppaoletti,2003jltsergey,2012prbbaggaley,
2011jbaggaley,2012pfzuccher,2014praallen,2017prfalberto,2019lan,2019galantucci}, while parameter $B_{1}$ and $B_{2}$ are dissipative parameter dependent.

\subsection{energy behavior and the general Magnus force}
\label{sube}

The corresponding Hamiltonian for the GPE model is
\begin{equation}
  H=\int d^{2}x(|\nabla\psi|^{2}+\frac{1}{2}|\psi|^{4}-|\psi|^{2}).
\end{equation}
Here we consider the energy of the uniform superfluid ($\psi=1$) to be zero, then the hamiltonian energy can be defined as
\begin{equation}
 E=\int d^{2}x(|\nabla\psi|^{2}+\frac{1}{2}|\psi|^{4}-|\psi|^{2}+\frac{1}{2}).
\end{equation}

\begin{figure}
		\centering
		\includegraphics[scale=0.45]{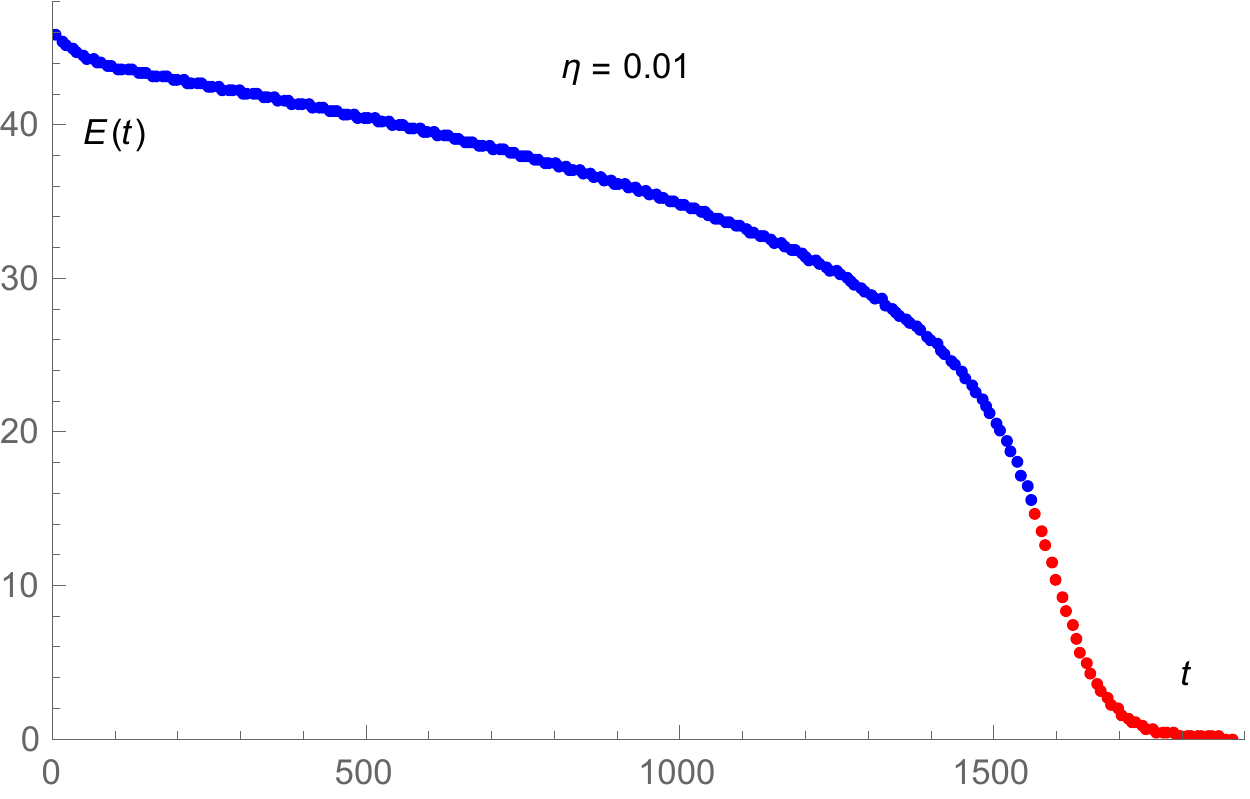}
        \includegraphics[scale=0.45]{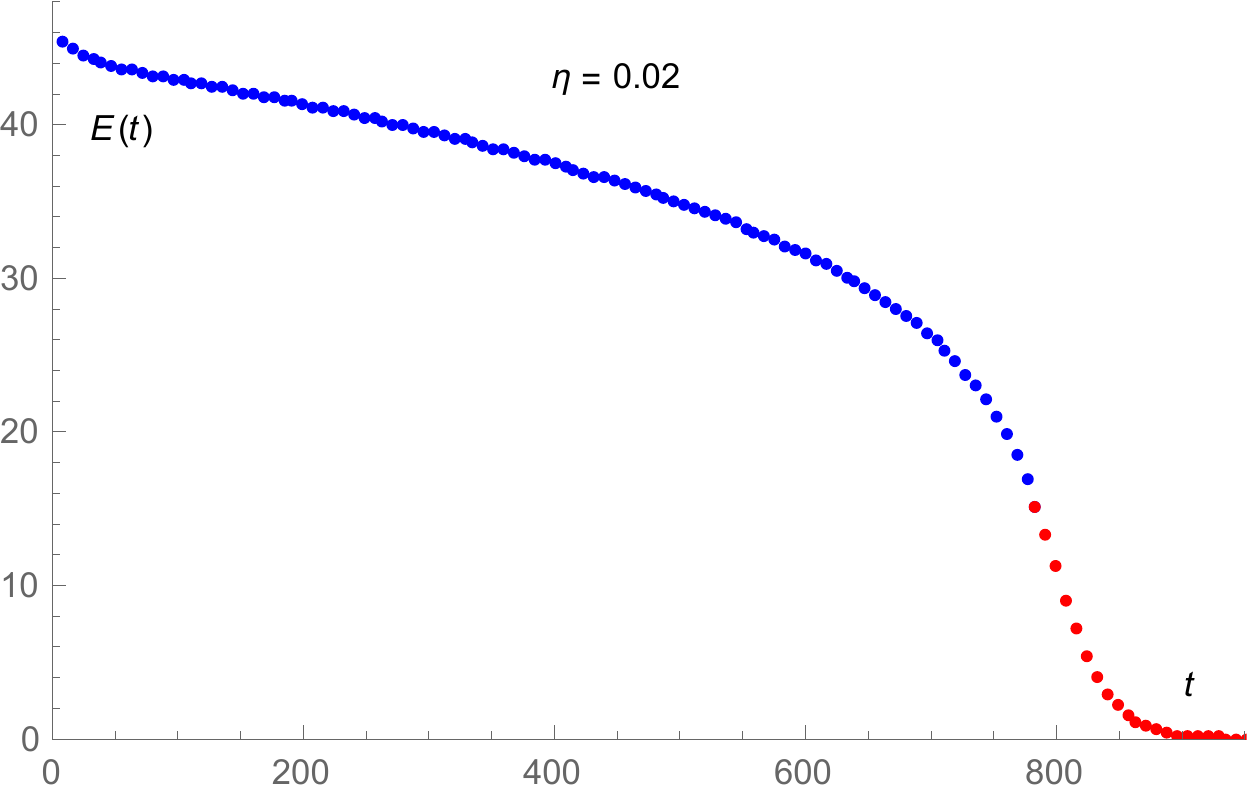}
        \includegraphics[scale=0.45]{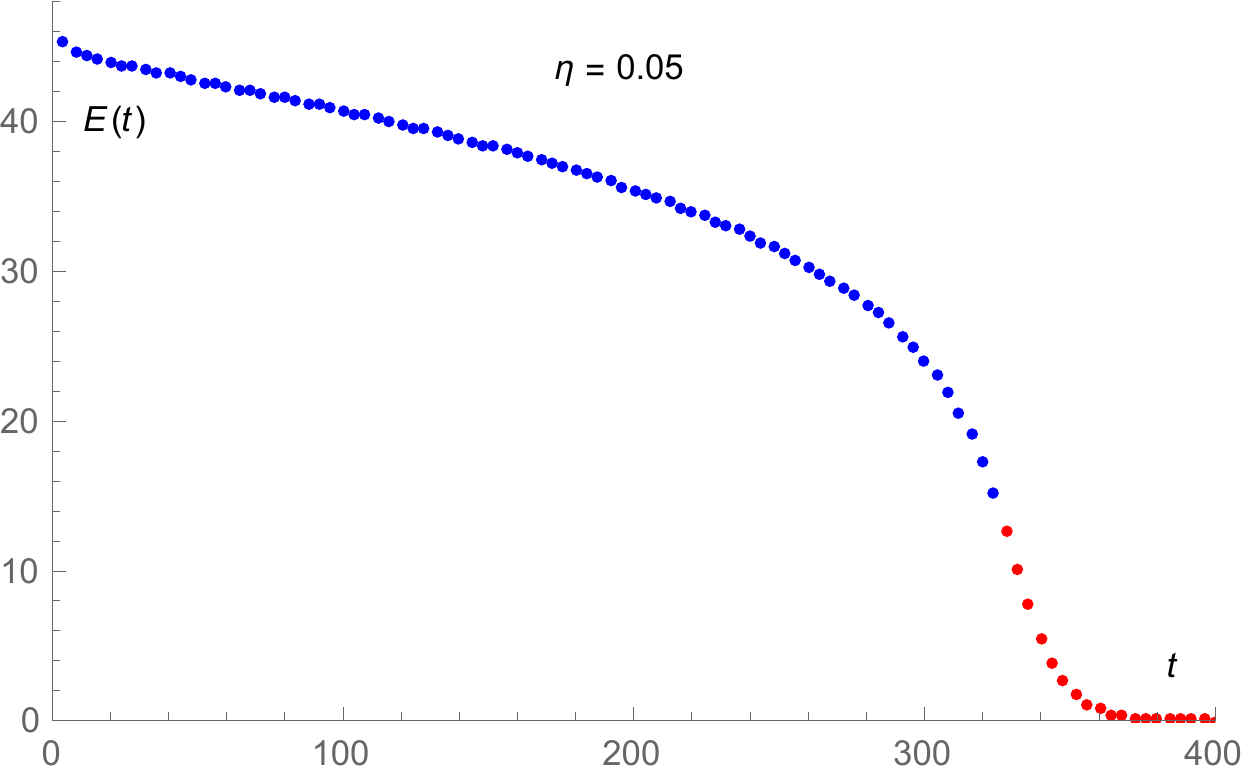}
        \includegraphics[scale=0.45]{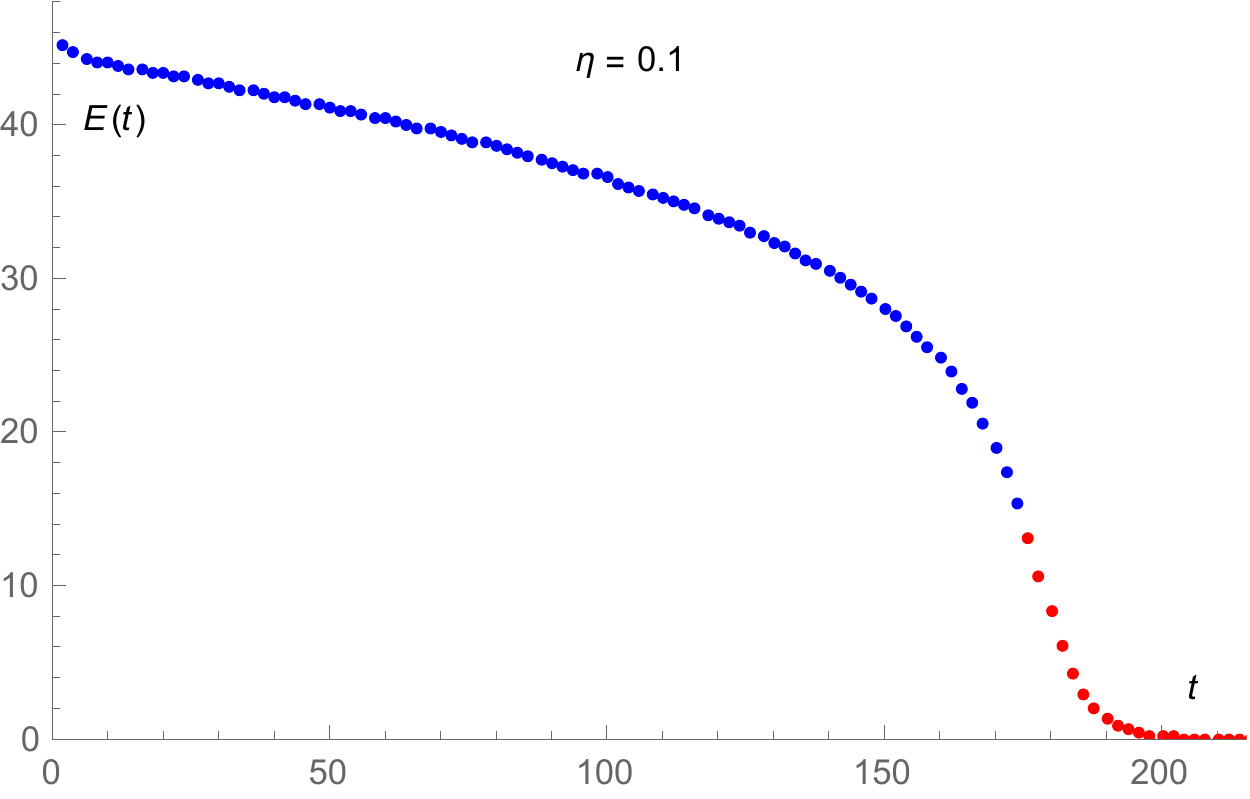}
	\caption{$E(t)$ vs. t. The blue dots denote vortex evolution stage and the red dots denote soliton and wave evolution stage. }
	\label{et}
\end{figure}

Fig.\ref{et} shows the energy evolution behavior for different dissipative parameters. The blue dots denote vortex evolution stage and the red dots denote soliton and wave evolution stage. During the whole process, system's energy is decreasing. When the separation distance of the vortices is large, the decreasing of energy is slow. When the separation distance of the vortices is small and soliton is formed, the decreasing of energy is dramatic. When the soliton becomes a wave, system's energy is very small and the decreasing of energy is very slow. The final state has zero energy which denotes the uniform superfluid. Energy of system possessing vortices is larger than that of uniform system. The four energy graphes are very similar, especially that the last blue dots seems to have the same energy, which make one wonder that the energy is configuration dependent.

\begin{figure}
		\centering
		\includegraphics[scale=0.8]{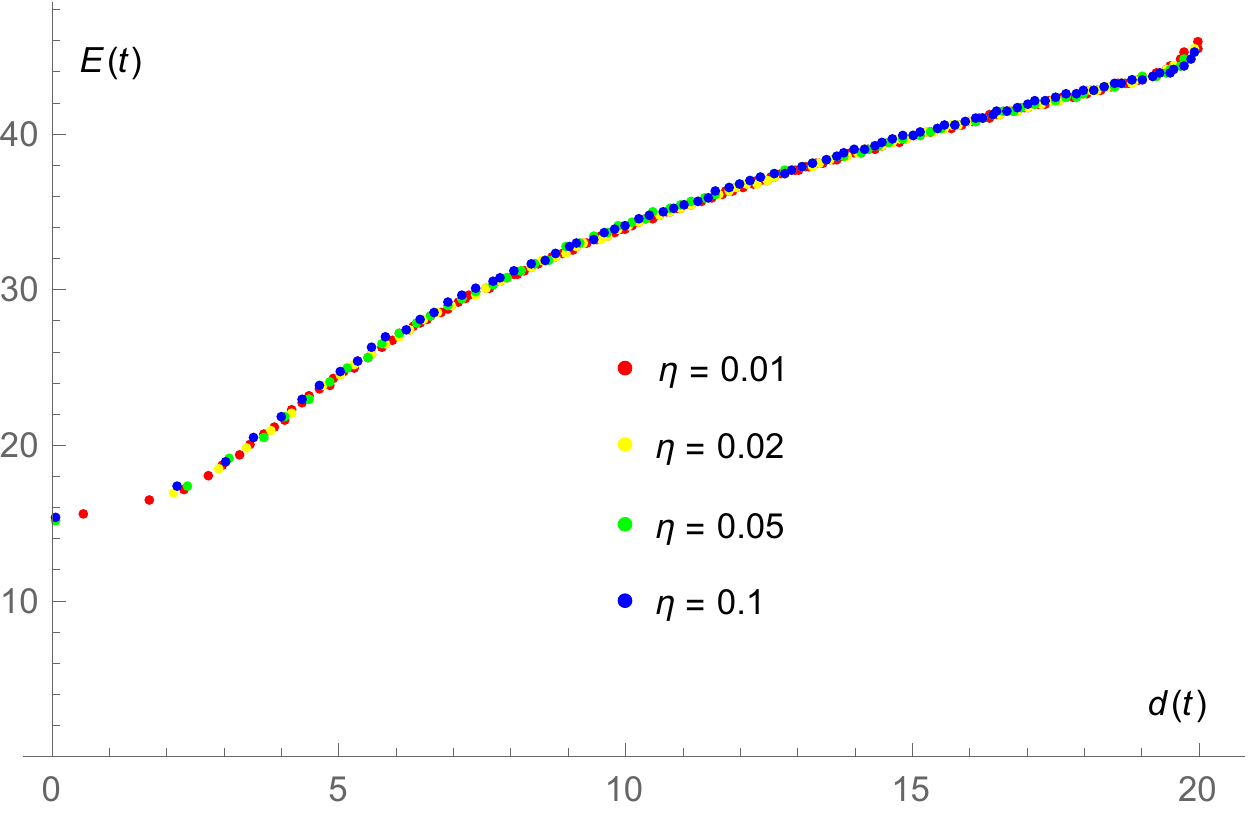}
	\caption{$E(t)$ vs. $d(t)$ for different dissipative parameters. As the four curves overlap, system's energy is configuration dependent.}
	\label{ed}
\end{figure}

Fig.\ref{ed} shows the $E(t)$ vs. $d(t)$ relation before the vortices annihilation. As the four curves overlap, system's energy is separation distance of vortices dependent. While the periodic system's configuration is determined by the separation distance of vortices. So one can conclude that system's energy is indeed configuration dependent. When the separation distance of the vortices is large, system's energy is large. When the separation distance of the vortices is small, system's energy is small. One can see that a newly formed soliton has large energy. The graphes' inflection point is around $(4.2,22.2)$.

When $d(t)>4.2$, the graphes are well fitted by
\begin{equation}
  E(d)=3.55+0.165 d+12.5 log_{e}(d).
\end{equation}
When $d(t)<4.2$, the graphes are well fitted by
\begin{equation}
  E(d)=15.3+0.407d-0.133d^{2}+0.189d^{3}-0.0209d^{4}.
\end{equation}
Although the system possesses both kinetic and potential energy, the total energy behaviors as a potential energy. Therefore, the derivation of energy versus distance can be regarded as a force.
When $d(t)>4.2$, the force is
\begin{eqnarray}\label{mgf}
  f(d)=-\frac{dE(d)}{dd}=-(0.165+\frac{12.5}{d}).
\end{eqnarray}
The minus denotes attractive force between the positive vortex and negative vortex. The force is actually the Magnus force. Note that the Magnus force is proportional to vorticity and drift velocity. In this large $d(t)$ case, vorticity can be considered as a constant and the Magnus force is proportional to drift velocity. The first term in the force formula is introduced by initial velocity which is introduced when we construct the periodic superfluid system. This initial velocity is small. The second term is introduced by the velocity field of the other vortex. Magnitude of the Magnus force decreases with the increase of $d(t)$ for large $d(t)$.
When $d(t)<4.2$, the force is
\begin{eqnarray}
  f(d)=-\frac{dE(d)}{dd}=-(0.407-0.266d+0.567d^{2}-0.0836d^{3}).
\end{eqnarray}
Although the Magnus force is not defined for small $d(t)$, we will still call it the Magnus force. Interestingly, this Magnus force increases with the increase of $d(t)$ for small $d(t)$. Therefore, despite the vorticity and drift velocity, the general Magnus force is also $d(t)$ dependent. As the general Magnus force have different behavior for $d(t)>4.2$ and $d(t)<4.2$, we define the vortex radius as $r=2.1$. One may call the annihilation process for $d(t)>4.2$ the approaching process and the annihilation process for $d(t)<4.2$ the colliding process.

\section{Conclusion and discussion}
\label{sec4}

A pair of vortices with winding number $\omega=\pm1$  are located in a two dimensional superfluid with periodic boundary. Then the vortices' annihilation processes are numerical simulated and investigated within the dissipative GPE model. First, the vortex configuration is obtained by solving the GPE in cylindrical coordination and a well fitting function is found. The vortex radius is roughly set to 2.1. Hence, the length scale of the vortex is 4.2 which separates the behavior of $d(t)$ and $E(d)$.

Second, vortices' annihilation processes  are analysed. The positive and negative vortex accelerate in the same positive x direction, and meantime they accelerate towards each other until they annihilate into a soliton and then a crescent-shaped shock wave. For different dissipative parameters $\eta$, the behaviors of vortices' trajectories, $x(t)$ and $y(t)$ are similar with different time scales $\tau\propto 1/\eta$. The acceleration in x direction is caused by $d(t)$ and the acceleration in y direction is caused by the Magnus force. As the acceleration in y direction increases with the dissipative parameters, the ``inertial mass" of vortex decreases with the increases of the dissipative parameter. For the behavior of separation distance between the vortices, a fitting function is found with an universal scaling exponent 1/2 which is same with the three dimensional cases. The fitting formula is accurate when $d(t)$ is large, e.g. $d(t)>4.2$.

Third, evolution of system's energy are shown during the whole stage. A newly formed soliton is found to possess large energy which decreases very fast. For different dissipative parameters, the energy behavior are similar. Then we plot the $E(t)-d(t)$ graphes which amazingly overlap. Their fitting functions are found for $d(t)>4.2$ and $d(t)<4.2$. Therefore, the general Magnus force is derived. The force has different behaviors for different $d(t)$. When $d(t)$ is large, as the drift velocity decreases with the increases of $d(t)$, the force will decreases. While, when $d(t)$ is small, the force increases with $d(t)$. In a word, the general Magnus force is the separation distance of vortices dependent.

In sec.\ref{subv}, we qualitatively rather than quantitatively discussed the behavior of vortices' trajectories, $x(t)$ and $y(t)$ in Fig.\ref{vorxy}, Fig.\ref{vorxt} and Fig.\ref{voryt}. That is because the well fitting functions for the graphes are not found, and poor fitting functions will make us get inaccurate vortices velocities and accelerations. Note that the background superfluid is static and the initial vortices velocities are small, as a result, the Magnus force is small and the annihilation process is slow. Thus it is difficult to find the well fitted functions for these graphes. If the background superfluid has a velocity, the problem may be solved. What's more, the annihilation processes of vortices for different background superfluid velocities are interesting in their own. Especially, in that cases, the Magnus force for large $d(t)$ should have the same formula as Eq.(\ref{mgf}).  Anyway, this is beyond the scope of this paper.

\begin{acknowledgments}
This research is supported by National Natural Science Foundation of China (Grant Nos.11847001), Department of Education of Guangdong Province, China (Grant Nos.2017KQNCX124) and the Lingnan Normal University Project ZL1931.
\end{acknowledgments}



\begin{thebibliography}{99}

\bibitem{hanninen2015}
R. H$\ddot{a}$nninen,
\emph{Kelvin waves from vortex reconnection in superfluid helium at low temperatures},
Phys. Rev. B 92, 184508 (2015).

\bibitem{serafini2017}
S. Serafini, L. Galantucci, E. Iseni, et al.
\emph{Vortex reconnections and rebounds in trapped atomic Bose--Einstein condensates},
Phys. Rev. X 7, 021031 (2017).

\bibitem{hannay2017}
J.H. Hannay,
\emph{Vortex reconnection rate, and loop birth rate, for a random wavefield},
J. Phys.A. Math. Theor. 50,16 (2017).

\bibitem{1988prbschwarz}
K. W. Schwarz,
Three-dimensional vortex dynamics in superfluid $^{4}$He: Homogeneous superfluid turbulence,
Phys. Rev. B 38, 2398-2417 (1988).

\bibitem{1994prlwaele}
A. D. Waele,  R. Aarts,
Route to vortex reconnection,
 Phys. Rev. Lett. 72, 482-485(1994).

\bibitem{2008PNASbewley}
   G. P. Bewley, M. S. Paoletti, K. R. Sreenivasan, and D. P. Lathrop,
   Characterization of reconnecting vortices in superfluid helium,
   Proceedings of the National Academy of Sciences 105, 13707-13710(2008).

\bibitem{2008pdnppaoletti}
   M. S. Paoletti, M. E. Fisher, and D. P. Lathrop,
   Reconnection dynamics for quantized vortices,
   Physica D: Nonlinear Phenomena 239, 1367-1377(2008).

\bibitem{2003jltsergey}
   S. Nazarenko, R. West,
   Analytical solution for nonlinear Schr$\ddot{o}$dinger vortex reconnection, J. Low. Temp. Phys. 132, 1-10(2003).

\bibitem{2011jbaggaley}
   R. Tebbs, A. J. Youd, and C. F. Barenghi,
    The approach to vortex reconnection,
     J. Low. Temp. Phys. 162, 314 (2011).

\bibitem{2012prbbaggaley}
   A. W. Baggaley, L. K. Sherwin, C. F. Barenghi, and Y. A. Sergeev,
    Thermally and mechanically driven quantum turbulence in helium II,
     Phys. Rev. B 86, 104501 (2012).

\bibitem{2012pfzuccher}
   S. Zuccher, M. Caliari, A. W. Baggaley, and C. F. Barenghi,
   Quantum vortex reconnections,
     Phys. Fluids 24, 125108(2012).

\bibitem{2014praallen}
   A. J. Allen, S. Zuccher, M. Caliari, N. P. Proukakis, N. G. Parker, and C. F. Barenghi,
   Vortex reconnections in atomic condensates at finite temperature,
     Phys. Rev. A 90, 013601 (2014).

\bibitem{2017prfalberto}
   A. Villois, G. Krstulovic, and D. Proment,
    (Non)-universality of vortex reconnections in superfluids,
     Phys. Rev. Fluids 2, 044701 (2017).

\bibitem{2019lan}
   S.Q. Lan, G.Q. Li, J.X. Mo and X.B. Xu,
   Attractive Interaction between Vortex and Anti-vortex in Holographic Superfluid,
     J. High Energ. Phys. 02, 122 (2019).

\bibitem{2019galantucci}
   L. Galantucci, A. W. Baggaley, N. G. Parker, and C. F. Barenghi,
   Crossover from interaction to driven regimes in quantum vortex reconnections,
     Proc. Natl. Acad. Sci. U.S.A. 116, 12204 (2019).

\bibitem {epgross1963}
    E. P. Gross,
   \emph{Hydrodynamics of a Superfluid Condensate},
    J. Math.Phys. 4, 195(1963).

\bibitem{pitaevskii1959}
L. P. Pitaevskii,
\emph{Phenomenological Theory of Superfluidity near the $\lambda$ Point},
Zh. Eksp. Teor. Fiz. 35, 408 (1959)[Sov. Phys. JETP 8, 282 (1959)].

\bibitem{lppitaevskii1961}
L. P. Pitaevskii,
\emph{Vortex Lines in an Imperfect Bose Gas},
Zh. Eksp. Teor. Fiz. 40, 646 (1961)
[Sov. Phys. JETP 13, 451 (1961)].

\bibitem {choi1998}
    S. Choi, S. A. Morgan, and K. Burnett,
   \emph{Phenomenological damping in trapped atomic Bose-Einstein condensates},
    Phys. Rev. A 57, 4057(1998).

\bibitem {tsubota2002}
    M. Tsubota, K. Kasamatsu, and M. Ueda,
   \emph{Vortex lattice formation in a rotating Bose-Einstein condensate},
    Phys. Rev. A 65, 023603(2002).

\bibitem{chesler2013holographic}
   P. M. Chesler, H. Liu, and A. Adams,
  Holographic vortex liquids and superfluid turbulence,
  Science 341, 368-372 (2013).

\bibitem{du2014holographic}
   Y. Q. Du, C. Niu, Y. Tian, and H. B. Zhang,
  Holographic thermal relaxation in superfluid turbulence,
  J. High Energy Phys. 12, 018 (2015).

\bibitem{2016jheplan}
   S. Q. Lan,  Y. Tian, and H. B. Zhang,
  Towards quantum turbulence in finite temperature Bose-Einstein condensates,
  J. High Energy Phys. 07, 092 (2016).

\bibitem{2016cqgguo}
   M. Y. Guo,  S. Q. Lan, Y. Tian, and H. B. Zhang,
  Note on zero temperature holographic superfluids,
  Class. Quant. Grav. 33, 127001 (2016).

\bibitem{ueda2010}
   M. Ueda ,
  Fundamentals and New Frontiers of Bose-Einstein Condensation,
  World Scientific Publishing,171-181 (2010).

\end{thebibliography}
\end{document}